\documentclass[aps,prd,12pt,%
  nofootinbib,superscriptaddress%
  ,eqsecnum
  ,preprint
]{revtex4}
\usepackage{graphicx}

\renewcommand{\thesection}{\arabic{section}}

\makeatletter
\renewcommand{\p@subsection}{}
\makeatother

\def\la{\langle}\def\ra{\rangle}
\def\bea{\begin{eqnarray}}
\def\eea{\end{eqnarray}}
\def\lsim{\mathrel{\rlap{\lower3pt\hbox{\hskip1pt$\sim$}}
     \raise1pt\hbox{$<$}}} %less than or approx. symbol
\def\gsim{\mathrel{\rlap{\lower3pt\hbox{\hskip1pt$\sim$}}
     \raise1pt\hbox{$>$}}} %greater than or approx. symbol

\def\chisu3{SU(3)_L\times SU(3)_R}

\def\calH{\mathcal H}

\def\calM{{\cal M}}

\begin{document}

\preprint{DPNU-03-26}
\title{Chiral Doubling of Heavy-Light Hadrons
\\ and the Vector Manifestation of Hidden Local Symmetry}

\vskip 1cm

\author{Masayasu Harada}
\affiliation{Department of Physics, Nagoya University,
Nagoya, 464-8602, Japan}
\author{Mannque Rho}
\affiliation{Service de Physique Th\'eorique, CEA Saclay, 91191
Gif-sur-Yvette, France} \affiliation{Department of Physics,
Hanyang University, Seoul 133-791, Korea}
\author{Chihiro Sasaki}
\affiliation{Department of Physics, Nagoya University,
Nagoya, 464-8602, Japan}

{\today}

\vskip 2cm

\begin{abstract}
Starting with a hidden local symmetry Lagrangian at the vector
manifestation (VM) fixed point that incorporates heavy-quark
symmetry and matching the bare theory to QCD, we calculate the
splitting of chiral doublers of heavy-light mesons proposed by
Nowak, Rho and Zahed~\cite{Nowak-Rho-Zahed:93} and Bardeen and
Hill~\cite{Bardeen-Hill:94}. We show, in the three-flavor chiral
limit, that the splitting is directly proportional to the
light-quark condensate $\la\bar{q}q\ra$ and comes out to be $\sim
\frac{1}{3} m_N$ where $m_N$ is the nucleon mass, implying that
the splitting vanishes in the chiral limit at the chiral
restoration point -- temperature $T_c$, density $n_c$ or number of
flavors $N_F^c$. The result turns out to be surprisingly simple
with the vector ($\rho$) meson playing the crucial role in quantum
corrections, pointing to the relevance of the VM to QCD in the way
chiral symmetry is manifested in hadronic matter. We also make
predictions on the hadronic decay processes of the excited heavy
(charm) -light mesons $\tilde{D}$.
\end{abstract}

\maketitle

%%%%%%%%%%%%%%%%%%%%%%%%%%%%%%%%%%%%%%%%%%%%%%%%%%%%%%%%%%%%%%%%%%%%
%%%%%%%%%%%%%%%%%%%%%%%%%%%%%%%%%%%%%%%%%%%%%%%%%%%%%%%%%%%%%%%%%%%%

\section{Introduction}

Based on the manifestation of chiral symmetry \`a la linear sigma
model, it was predicted a decade
ago~\cite{Nowak-Rho-Zahed:93,Bardeen-Hill:94} that the mass
splitting $\Delta M$ between the $\calM (0^-, 1^-)$ and
$\tilde{\calM}(0^+,1^+)$ mesons where $\calM$ denotes a
heavy-light meson consisting of heavy qaurk $Q$ and light
antiquark $\bar{q}$ should be of the size of the constituent quark
mass. Recently, BaBar~\cite{BABAR}, CLEO~\cite{CLEO} and
subsequently the Belle collaboration~\cite{Belle}, discovered new
$D$ mesons with $Q=c$, $c$ being charm quark, which most likely
have spin-party $0^+$ and $1^+$ and the mass difference to the
$D(0^-, 1^-)$ is in fair agreement with the prediction of
\cite{Nowak-Rho-Zahed:93,Bardeen-Hill:94}. In a recent article,
Nowak, Rho and Zahed~\cite{Nowak-Rho-Zahed:03} proposed that the
splitting of $\calM$ and $\tilde{\calM}$ mesons could carry direct
information on the property of chiral symmetry at some critical
density or temperature at which the symmetry is
restored~\footnote{In what follows, unless otherwise specified, we
will refer to heavy-light mesons generically as $D$ but the
arguments should apply better to heavier-quark mesons. Numerical
estimates will however be made solely for the (open charm) $D$
mesons.}. In this paper we pick up this idea and make a first step
in consolidating the proposal of Ref.~\cite{Nowak-Rho-Zahed:03}.
In doing this, we shall take the reverse direction: Instead of
starting with a Lagrangian defined in the chiral-symmetry broken
phase and then driving the system to the chiral symmetry
restoration point by an external disturbance, we will start from
an assumed structure of chiral symmetry at its restoration point
and then make a prediction as to what happens to the splitting in
the broken phase. We find that the splitting is directly
proportional to the light-quark condensate and comes out to be of
the size of the constituent quark mass consistent with the
prediction of Refs.~\cite{Nowak-Rho-Zahed:93,Bardeen-Hill:94}. We
shall associate this result as giving a link between the assumed
structure of the chiral restoration point and the broken phase.

Our procedure is based on the result that
the effective field theory (EFT) implementing hidden local symmetry
(HLS)~\cite{BKUYY,BKY},
when matched to QCD at a suitable matching scale
$\Lambda_M$, represents QCD up to the matching
scale~\cite{HY:WM,HY:PRep}
and
the ``vector manifestation" (VM)~\cite{HYb,HY:PRep}
is realized
in the chiral limit at the chiral
restoration point generically denoted $C_\chi$
(critical temperature $T_c$~\cite{HS:VMT}
or density $n_c$~\cite{HKR}
or number of flavors $N_F^c$~\cite{HYb}).
In the HLS theory consisting of pions and vector
mesons, the VM is characterized by the existence of a fixed point
called VM fixed point at which the HLS gauge coupling constant $g$
and the vector meson mass $m_V$ vanish with the longitudinal
components of the vector mesons joining in the multiplet with the
pions and the pion decay constant $f_\pi\rightarrow 0$. In this
theory -- referred to in short as HLS/VM, the system flows
uniquely to the fixed point as one approaches $C_\chi$ from below.
The VM fixed point implies that light vector meson masses vanish
proportionally to the quark condensate $\la\bar{q}q\ra$ -- the
order parameter of chiral symmetry -- as one approaches $C_\chi$,
supporting the scenario suggested in BR scaling~\cite{BR91}.
We
assume that the heavy-light hadrons are described by a VM-fixed
point theory at $C_\chi$ and by introducing the simplest form of
the VM breaking terms, we compute the mass splitting of the chiral
doublers in matter-free space in terms of the quantities that
figure in the QCD correlators~\footnote{
  Introducing vector mesons in the light-quark sector
  of heavy-light mesons was considered in
  Ref.~\cite{Nowak-Rho-Zahed:93} but without
  the matching to QCD and hence without the VM fixed point.
}.

Before going into our main theme, we should note that the presence
of light vector mesons near the VM fixed point makes certain
predictions that are basically different from the standard
scenario in which the only relevant (hadronic) degrees of freedom
near the critical point are the pions (and a light scalar). For
instance, the HLS/VM~\cite{HY:PRep} predicts that the pion
velocity approaches the speed of light as $T\rightarrow
T_c$~\cite{HKRS} in a stark contrast to the standard picture where
the pion velocity goes to zero~\cite{sonstephanov}. Whether or not
the light vectors do actually figure importantly in the vicinity
of the chiral phase transition should ultimately be checked by
lattice calculations. At the moment, there is no clear evidence
either for or against the VM scenario: What is needed but not yet
available is measurement of dynamical correlation functions. The
forthcoming ``maximum entropy method (MEM)"
analysis~\cite{hatsuda} for excitations just {\it below} $T_c$
might shed light on this important issue. In this paper, we shall
simply assume that the chiral restoration is described by HLS/VM
and ask whether this assumption is consistent with the splitting
observed by BaBar, CLEO and Belle.

This paper is organized as follows: In section~\ref{sec:Lag}, we
write down the EFT Lagrangian that defines our approach. In
section~\ref{sec:MOPE} we perform the matching to determine the
bare parameters of the EFT Lagrangian. Section~\ref{sec:QC} is
devoted to computing the quantum correction to the mass splitting
and obtaining the renormalization group equation for the parameter
expressing the splitting. In section~\ref{sec:DMD}, we give a
semi-quantitative estimate of the value of the mass splitting. To
see whether or not our scenario based on the VM differs from that
\`a la linear sigma model, we study the consequences of our
scenario on the hadronic decay processes of the open charm
$\tilde{D}$ meson in section~\ref{sec:HDM}. We give a brief
summary and discussions in section~\ref{sec:SD}.

%%%%%%%%%%%%%%%%%%%%%%%%%%%%%%%%%%%%%%%%%%%%%%%%%%%%%%%%%%%%%%%%%%%%%
%%%%%%%%%%%%%%%%%%%%%%%%%%%%%%%%%%%%%%%%%%%%%%%%%%%%%%%%%%%%%%%%%%%%%

\section{\label{sec:Lag} Lagrangian}

In this section we give our reasoning that leads to the Lagrangian
that defines our approach.
Here we construct the Lagrangian using the approximate chiral
$SU(3)_{\rm L}\times SU(3)_{\rm R}$
symmetry in the light-quark sector and the
heavy quark symmetry in the heavy-quark sector.
We will start
from the Lagrangian given at the vector manifestation
(VM) fixed point.
We first describe
how to construct the fixed point Lagrangian based on the VM.
Then,
we account for
the effect of spontaneous chiral symmetry breaking by adding a
$bare$ parameter for the mass splitting in the heavy sector and
including the deviation of the HLS parameters from the values
at the VM fixed point.
The explicit form of the Lagrangian so
constructed is shown in subsection~\ref{ssec:Lag}.

\subsection{The fixed point Lagrangian}
\label{ssec:FPL}

To define the notations, we briefly review the model based on the
hidden local symmetry (HLS)~\cite{BKUYY,BKY}. The HLS
model~\footnote{In the modern interpretation~\cite{georgi},
implementing HLS in the chiral Lagrangian can be associated with
the ``ultraviolet completion" to the fundamental theory of strong
interactions, i.e., QCD. The matching to QCD at a matching scale
is therefore a crucial ingredient of the approach.} is based on
the $G_{\rm{global}} \times H_{\rm{local}}$ symmetry, where
$G=SU(3)_{\rm L} \times SU(3)_{\rm R}$ is the chiral symmetry and
$H=SU(3)_{\rm V}$ is the HLS. The basic quantities are the HLS
gauge boson and two matrix valued variables $\xi_{\rm L}(x)$ and
$\xi_{\rm R}(x)$ which transform as
 \begin{equation}
  \xi_{\rm L,R}(x) \to \xi^{\prime}_{\rm L,R}(x)
  =h(x)\cdot \xi_{\rm L,R}(x)\cdot g^{\dagger}_{\rm L,R}\ ,
 \end{equation}
where $h(x)\in H_{\rm{local}}\ \mbox{and}\ g_{L,R}\in
[SU(3)_{\rm L,R}]_{\rm{global}}$.
These variables are parameterized as
 \begin{equation}
  \xi_{\rm L,R}(x)
  =e^{i\sigma (x)/{F_\sigma}}e^{\mp i\pi (x)/{F_\pi}}\ ,
 \end{equation}
where $\pi = \pi^a T_a$ denotes the pseudoscalar Nambu-Goldstone
bosons associated with the spontaneous symmetry breaking of
$G_{\rm{global}}$ chiral symmetry, and $\sigma = \sigma^a T_a$
denotes the Nambu-Goldstone bosons associated with the spontaneous
breaking of $H_{\rm{local}}$. This $\sigma$ is eaten up by the HLS
gauge boson becoming massive through the Higgs mechanism. $F_\pi \
\mbox{and}\ F_\sigma$ are the decay constants of the associated
particles. The phenomenologically important parameter $a$ is
defined by the ratio
 \begin{equation}
  a = \frac{{F_\sigma}^2}{{F_\pi}^2}\ .
 \end{equation}
The covariant derivatives of $\xi_{\rm L,R}$ are given by
\begin{eqnarray}
 D_\mu \xi_{\rm L} &=& \partial_\mu\xi_{\rm L} - iV_\mu \xi_{\rm L}
 \ ,
 \nonumber\\
 D_\mu \xi_{\rm R} &=& \partial_\mu\xi_{\rm R} - iV_\mu \xi_{\rm R}
 \ ,
\end{eqnarray}
where $V_\mu$ is the gauge field of $H_{\rm{local}}$.

The basic quantities in constructing the Lagrangian are the
following two 1-forms:
\begin{eqnarray}
  \hat{\alpha}_{\parallel\mu} =
  \frac{1}{2i} \left(
    D_\mu \xi_{\rm R} \cdot \xi_{\rm R}^\dag +
    D_\mu \xi_{\rm L} \cdot \xi_{\rm L}^\dag
  \right)
\ ,
\nonumber\\
  \hat{\alpha}_{\perp\mu} =
  \frac{1}{2i} \left(
    D_\mu \xi_{\rm R} \cdot \xi_{\rm R}^\dag -
    D_\mu \xi_{\rm L} \cdot \xi_{\rm L}^\dag
  \right)
\ .
\label{alpha}
\end{eqnarray}
They transform as
\begin{eqnarray}
  \hat{\alpha}_{\perp,\parallel}^{\mu} \rightarrow
  h(x) \cdot \hat{\alpha}_{\perp,\parallel}^{\mu} \cdot h^\dag(x)
\ .
\label{transform}
\end{eqnarray}
When HLS is gauge-fixed to the unitary gauge, $\sigma=0$,
$\xi_{\rm L}$ and $\xi_{\rm R}$ are related with each other by
\begin{equation}
\xi_{\rm L}^\dag = \xi_{\rm R} \equiv
\xi = e^{i\pi/F_\pi} \ .
\end{equation}
This unitary gauge is not preserved under the $G_{\rm global}$
transformation, which in general has the following form
\begin{eqnarray}
 G_{\rm global} \ : \  \xi \ \rightarrow \ \xi^\prime
&=&
 \xi \cdot g_{\rm R}^\dag = g_{\rm L} \cdot \xi
\nonumber\\
&=&
 \exp \left[i \sigma^\prime (\pi,g_{\rm R},g_{\rm L})/F_\sigma\right]
 \exp \left[ i\pi^\prime/F_\pi\right]
\nonumber\\
&=&
 \exp \left[ i\pi^\prime/F_\pi\right]
 \exp \left[-i\sigma^\prime (\pi,g_{\rm R},g_{\rm L})/F_\sigma\right]
\ .
\end{eqnarray}
The unwanted factor
$\exp \left[i \sigma^\prime (\pi,g_{\rm R},g_{\rm L})/F_\sigma\right]$
can be eliminated if we simultaneously perform the
$H_{\rm local}$ gauge transformation with
\begin{equation}
H_{\rm local} \ : \
h = \exp
  \left[i \sigma^\prime (\pi,g_{\rm R},g_{\rm L})/F_\sigma\right]
\equiv h(\pi,g_{\rm R},g_{\rm L})
\ .
\end{equation}
Then the system has a global symmetry
$G=SU(3)_{\rm L} \times SU(3)_{\rm R}$
under the following combined transformation:
\begin{equation}
G \ : \
 \xi \ \rightarrow\
 h(\pi,g_{\rm R},g_{\rm L}) \cdot \xi \cdot g_{\rm R}^\dag
 = g_{\rm L} \cdot \xi \cdot h(\pi,g_{\rm R},g_{\rm L})
\ .
\end{equation}
Under this transformation the HLS gauge boson field $V_\mu$
in the unitary gauge transforms as
\begin{equation}
G \ : \
 V_\mu \ \rightarrow\
 h(\pi,g_{\rm R},g_{\rm L}) \cdot  V_\mu
 \cdot h^\dag(\pi,g_{\rm R},g_{\rm L})
 - i \partial h(\pi,g_{\rm R},g_{\rm L}) \cdot
  h^\dag(\pi,g_{\rm R},g_{\rm L})
\ ,
\end{equation}
which is nothing but the transformation property of
Weinberg's ``$\rho$ meson''~\cite{Wei:68}.
The two 1-forms $\hat{\alpha}_{\parallel}^{\mu}$ and
$\hat{\alpha}_{\perp}^{\mu}$ transform as
\begin{eqnarray}
  \hat{\alpha}_{\perp,\parallel}^{\mu} \rightarrow
  h(\pi,g_{\rm R},g_{\rm L}) \cdot
  \hat{\alpha}_{\perp,\parallel}^{\mu} \cdot
  h^\dag(\pi,g_{\rm R},g_{\rm L})
\ .
\end{eqnarray}
Then, we can regard these 1-forms
as the fields belonging to
the chiral representations
$(1,8) + (8,1)$ and $(1,8) - (8,1)$
under $SU(3)_{\rm L} \times SU(3)_{\rm R}$.

Let us next consider the VM at the point at which chiral symmetry
is restored (in the chiral limit). At the VM at its fixed point
characterized by $(g,a)=(0,1)$, the two 1-forms become
\begin{eqnarray}
  {\alpha}_{\parallel\mu} =
  \frac{1}{2i} \left(
    {\partial}_\mu\xi_{\rm R} \cdot \xi_{\rm R}^\dag +
    {\partial}_\mu \xi_{\rm L} \cdot \xi_{\rm L}^\dag
  \right)
\ ,
\nonumber\\
  {\alpha}_{\perp\mu} =
  \frac{1}{2i} \left(
    {\partial}_\mu \xi_{\rm R} \cdot \xi_{\rm R}^\dag -
    {\partial}_\mu \xi_{\rm L} \cdot \xi_{\rm L}^\dag
  \right)
\ .
\end{eqnarray}
Note that the above ${\alpha}_{\parallel\mu}$ and
${\alpha}_{\perp\mu}$ do not
contain the HLS gauge field since the gauge coupling $g$ vanishes
at the VM fixed point. It is convenient to define the (L,R)
1-forms:
\begin{eqnarray}
{\alpha}_{R\mu} &=&
  {\alpha}_{\parallel\mu} + {\alpha}_{\perp\mu}
  = \frac{1}{i} \partial_\mu \xi_{\rm R} \cdot \xi^\dag_{\rm R}
\ ,
\nonumber\\
{\alpha}_{L\mu} &=&
  {\alpha}_{\parallel\mu} - {\alpha}_{\perp\mu}
  = \frac{1}{i} \partial_\mu \xi_{\rm L} \cdot \xi^\dag_{\rm L}
\ ,
\label{def:1forms}
\end{eqnarray}
which can be regarded as belonging to the chiral representation
$(1,8)$ and $(8,1)$, respectively, transforming under chiral
$SU(3)_{\rm L} \times SU(3)_{\rm R}$ as
\begin{eqnarray}
{\alpha}_{R\mu} \rightarrow
g_{\rm R} {\alpha}_{R\mu} g_{\rm R}^\dag
\ ,
\nonumber\\
{\alpha}_{L\mu} \rightarrow
g_{\rm L} {\alpha}_{L\mu} g_{\rm L}^\dag
\ .
\end{eqnarray}
By using these 1-forms, the HLS Lagrangian at the VM fixed point
can be written as~\cite{HY:PRep}
\begin{eqnarray}
{\mathcal L}_{\rm light}^{\ast}
=
\frac{1}{2} F_\pi^2 \,\mbox{tr}\left[
  {\alpha}_{R\mu} {\alpha}_{R}^\mu
\right]
+
\frac{1}{2} F_\pi^2 \,\mbox{tr}\left[
  {\alpha}_{L\mu} {\alpha}_{L}^\mu
\right]
\ ,
\label{light fixed Lag}
\end{eqnarray}
where the $*$ affixed to the
Lagrangian denotes that it is a fixed-point Lagrangian,
and $F_\pi$ denotes the bare pion decay constant.
Note that the physical pion decay
constant $f_\pi$ vanishes at the VM fixed point by the quadratic
divergence although the bare one is non-zero~\cite{HY:PRep}. It
should be stressed that the above fixed point Lagrangian is
approached only as a limit of chiral symmetry
restoration~\cite{HY:PRep}.

Next we consider the fixed-point Lagrangian of the heavy meson
sector at the chiral restoration point identified with the VM
fixed point. Let us introduce two heavy-meson fields
${\mathcal H}_R$ and
${\mathcal H}_L$ transforming under chiral
$\mbox{SU}(3)_{\rm R} \times\mbox{SU}(3)_{\rm L}$ as
\begin{equation}
{\mathcal H}_R \rightarrow {\mathcal H}_R \, g_{\rm R}^\dag \ ,
\quad
{\mathcal H}_L \rightarrow {\mathcal H}_L \, g_{\rm L}^\dag \ .
\end{equation}
By using these fields together with
the light-meson 1-forms $\alpha_{L,R}^\mu$,
the fixed point Lagrangian of the heavy
mesons
is expressed as~\footnote{
  We assign the right chirality to ${\mathcal H}_R$, and the left
  chirality to ${\mathcal H}_L$. Then the interaction term has the
  right and left projection operators. Note that the insertion of
  $(1\pm\gamma_5)$
  to kinetic and mass termes does not cause any difference.
}
\begin{eqnarray}
{\mathcal L}_{\rm heavy}^{\ast}
&=&
- \mbox{tr} \left[
  {\mathcal H}_{R} i v_\mu \partial^\mu \bar{\mathcal H}_{R}
\right]
- \mbox{tr} \left[
  {\mathcal H}_{L} i v_\mu \partial^\mu \bar{\mathcal H}_{L}
\right]
\nonumber\\
&&
{} + m_0 \,\mbox{tr} \left[
  {\mathcal H}_R \bar{\mathcal H}_R
  + {\mathcal H}_L \bar{\mathcal H}_L
\right]
\nonumber\\
&&
{} + 2k \,\mbox{tr}\left[
  {\mathcal H}_R {\alpha}_{R\mu} \gamma^\mu \frac{1+\gamma_5}{2}
  \bar{\mathcal H}_R
  +
  {\mathcal H}_L {\alpha}_{L\mu} \gamma^\mu \frac{1-\gamma_5}{2}
  \bar{\mathcal H}_L
\right]
\ ,
\label{heavy fixed Lag}
\end{eqnarray}
where $v_\mu$ is the velocity of heavy meson, $m_0$ represents the
mass generated by the interaction between heavy quark and the
``pion cloud" surrounding the heavy quark, and $k$ is a real
constant to be determined.

\subsection{Effects of spontaneous chiral symmetry breaking}
\label{ssec:TESCSB}

Next we consider what happens in the broken phase of chiral
symmetry.
In the real world at low temperature and low density, chiral
symmetry is spontaneously broken by the non-vanishing quark
condensate. In the scenario of chiral-symmetry manifestation \`a
la linear sigma model, the effect of spontaneous chiral symmetry
breaking is expressed by the vacuum expectation value
(VEV) of the scalar fields.
In the VM,
on the other hand, it is signalled by the HLS
Lagrangian departing from the VM fixed point:
There the gauge
coupling constant $g \neq 0$~\footnote{
  Actually, near the chiral restoration point, the
  Wilsonian matching between HLS and QCD dictates~\cite{HY:PRep}
  that (in the chiral limit) the HLS gauge coupling be proportional
  to the quark condensate:
  $g \sim \langle \bar{q} q \rangle$.
}
and we have the kinetic term of the
HLS gauge bosons ${\mathcal L}_{\rho {\rm kin}} = - \frac{1}{2}
\mbox{tr} \left[
  \rho_{\mu\nu} \rho^{\mu\nu}
\right]$.
The derivatives in the HLS 1-forms become the covariant
derivatives and then $\alpha_{L\mu}$ and $\alpha_{R\mu}$ are
covariantized:
\begin{eqnarray}
 &\partial_\mu \to D_\mu = \partial_\mu - ig\rho_\mu,&
\nonumber\\
 &\alpha_{R\mu}
  \to \hat{\alpha}_{R\mu} = \alpha_{R\mu} - g\rho_\mu,&
\nonumber\\
 &\alpha_{L\mu}
  \to \hat{\alpha}_{L\mu} = \alpha_{L\mu} - g\rho_\mu.&
\label{covariant}
\end{eqnarray}
These 1-forms transform as $\hat{\alpha}_{R(L)\mu} \to h\,
\hat{\alpha}_{R(L)\mu}h^\dagger$ with
$h \in [SU(3)_{\rm V}]_{\rm local}$
as shown in Eq.~(\ref{transform}).

Although $a=1$ at the VM fixed point, generally $a \neq 1$ in the
broken phase. We therefore expect to have a term of the form
$\frac{1}{2}(a-1)F_\pi^2 \mbox{tr}[\hat{\alpha}_{L\mu}
\hat{\alpha}_R^\mu]$. Thus the Lagrangian for the light mesons
takes the following form:
\begin{eqnarray}
 {\cal L}_{\rm light}
 &=& \frac{a+1}{4}F_\pi^2 \mbox{tr}
  [ \hat{\alpha}_{R\mu}\hat{\alpha}_R^\mu
    {}+ \hat{\alpha}_{L\mu}\hat{\alpha}_L^\mu ]
\nonumber\\
 &&\qquad
  {}+ \frac{a-1}{2}F_\pi^2 \mbox{tr}
   [ \hat{\alpha}_{R\mu}\hat{\alpha}_L^\mu ]
  {}+ {\cal L}_{\rho{\rm kin}}.
\end{eqnarray}
By using $\hat{\alpha}_{\parallel\mu}$ and $\hat{\alpha}_{\perp\mu}$
given in Eq.~(\ref{alpha}), this Lagrangian is rewritten as
\begin{equation}
{\cal L}_{\rm light} =
  F_\pi^2 \mbox{tr}[ \hat{\alpha}_{\perp\mu}\hat{\alpha}_\perp^\mu ]
 {}+ F_\sigma^2 \mbox{tr}[ \hat{\alpha}_{\parallel\mu}
     \hat{\alpha}_\parallel^\mu ]
 {}+ {\cal L}_{\rho {\rm kin}},
\end{equation}
which is nothing but the general HLS Lagrangian.

We next consider the spontaneous breaking of chiral symmetry in
the heavy-meson sector. One of the most important effects of the
symmetry breaking is to generate the mass splitting between the
odd parity multiplet and the even parity
multiplet~\cite{Nowak-Rho-Zahed:93}. This effect can be
represented by the Lagrangian of the form:
\begin{eqnarray}
{\mathcal L}_{\chi{\rm{SB}}}
= \frac{1}{2}
\Delta M \,\mbox{tr}\left[
  {\mathcal H}_{L} \bar{{\mathcal H}}_R +
  {\mathcal H}_R \bar{{\mathcal H}}_L
\right]
\ ,
\label{bare Delta M}
\end{eqnarray}
where ${\mathcal H}_{R(L)}$ transforms under the HLS as
${\mathcal H}_{R(L)} \rightarrow {\mathcal H}_{R(L)}\,h^\dag$.
Here
$\Delta M$ is the $bare$ parameter corresponding to the mass
splitting between the two multiplets. An important point of our
work is that the bare $\Delta M$ can be determined by matching the
EFT with QCD as we will show in
section~\ref{sec:DMD}:
The
matching actually shows that $\Delta M$ is proportional to the
quark condensate:
\begin{equation}
\Delta M \sim \langle \bar{q} q \rangle.
\label{bare mass diff 0}
\end{equation}

\subsection{Lagrangian in parity eigenfields}
\label{ssec:Lag}

In order to compute the mass splitting between $\calM$ and
$\tilde{\calM}$, it is convenient to go to the corresponding
fields in parity eigenstate, $H$ (odd-parity) and $G$
(even-parity) as defined, e.g., in Ref.~\cite{Nowak-Rho-Zahed:03};
\begin{eqnarray}
{\mathcal H}_R &=&
  \frac{1}{\sqrt{2}} \left[ G - i H \gamma_5 \right] \ ,
\nonumber\\
{\mathcal H}_L &=&
  \frac{1}{\sqrt{2}} \left[ G + i H \gamma_5 \right] \ .
\end{eqnarray}
Here, the pseudoscalar meson $P$ and the vector meson $P_\mu^{\ast}$
are included in the $H$ field as
\begin{eqnarray}
H &=& \frac{1+ v_\mu \gamma^\mu }{2} \left[
  i \gamma_5 P + \gamma^\mu P_\mu^{\ast}
\right]
\ ,
\label{H}
\end{eqnarray}
and the scalar meson $Q^{\ast}$ and the axial-vector meson
$Q_\mu$ are in $G$ as
\begin{eqnarray}
G &=& \frac{1+ v_\mu \gamma^\mu }{2} \left[
  Q^\ast - i \gamma^\mu \gamma_5 Q_\mu
\right]
\ .
\label{G}
\end{eqnarray}
In terms of the $H$ and $G$ fields, the heavy-meson Lagrangian
{\it off} the VM fixed point is of the form
\begin{equation}
 {\cal L}_{\rm heavy} = {\cal L}_{\rm kin} + {\cal L}_{\rm int}
 \ ,
\end{equation}
with
\begin{eqnarray}
{\mathcal L}_{\rm kin} &=&
 \mbox{tr} \left[ H \,
   \left( i v_\mu D^\mu - M_H \right)
 \bar{H} \right]
 -
 \mbox{tr} \left[ G \,
   \left( i v_\mu D^\mu - M_G \right)
 \bar{G} \right]
\ ,
\\
{\cal L}_{\rm int}
 &=& k\, \Biggl[ \mbox{tr}[ H \gamma_\mu \gamma_5
     \hat{\alpha}_\perp^\mu \bar{H}]
 {}- \mbox{tr}[ H v_\mu
     \hat{\alpha}_\parallel^\mu \bar{H}]
\nonumber\\
 &&\qquad
 {}+ \mbox{tr}[ G \gamma_\mu \gamma_5
     \hat{\alpha}_\perp^\mu \bar{G}]
 {}+ \mbox{tr}[ G v_\mu
     \hat{\alpha}_\parallel^\mu \bar{G}]
\nonumber\\
 &&\qquad
 {}- i\mbox{tr}[ G \hat{\alpha}_{\perp\mu} \gamma^\mu
        \gamma_5 \bar{H}]
 {}+ i\mbox{tr}[ H \hat{\alpha}_{\perp\mu} \gamma^\mu
        \gamma_5 \bar{G}]
\nonumber\\
 &&\qquad
 {}- i\mbox{tr}[ G \hat{\alpha}_{\parallel\mu}
      \gamma^\mu \bar{H}]
 {}+ i\mbox{tr}[ H \hat{\alpha}_{\parallel\mu}
      \gamma^\mu \bar{G}] \Biggr],
\label{Lag:component}
\end{eqnarray}
where
the covariant derivatives acting on $\bar{H}$ and $\bar{G}$
are defined as
\begin{equation}
D_\mu \bar{H} = \left( \partial_\mu - i g \rho_\mu \right) \bar{H}
\ , \quad
D_\mu \bar{G} = \left( \partial_\mu - i g \rho_\mu \right) \bar{G}
\ .
\end{equation}
In the above expression, $M_H$ and $M_G$ denote the masses of
the parity-odd multiplet $H$ and the parity-even multiplet $G$,
respectively.
They are related to $m_0$ and $\Delta M$ as
\begin{eqnarray}
M_H &=& - m_0 - \frac{1}{2} \Delta M \ , \nonumber\\
M_G &=& - m_0 + \frac{1}{2} \Delta M \ .
\end{eqnarray}
The mass splitting between $G$ and $H$ is therefore given by
\begin{equation}
M_G - M_H = \Delta M \ .
\end{equation}

%%%%%%%%%%%%%%%%%%%%%%%%%%%%%%%%%%%%%%%%%%%%%%%%%%%%%%%%%%%%%%%%%%%%%
%%%%%%%%%%%%%%%%%%%%%%%%%%%%%%%%%%%%%%%%%%%%%%%%%%%%%%%%%%%%%%%%%%%%%
\section{Matching to the Operator Product Expansion}
\label{sec:MOPE}

The bare parameter $\Delta M_{{\rm bare}}$ which carries
information on QCD should be determined by matching the EFT
correltators to QCD ones. We are concerned with the pseudoscalar
correlator $G_P$ and the scalar correlator $G_S$. In the EFT
sector, the correlators at the matching scale are of the
form~\footnote{Here and in the rest of the paper, the heavy meson
is denoted $D$ with the open charm heavy meson in mind. However
the arguments (except for the numerical values) are generic for
all heavy mesons $\calM$.}
\begin{eqnarray}
 &&
 G_P(Q^2) = \frac{F_D^2 M_D^4}{M_D^2 + Q^2},
\nonumber\\
 &&
 G_S(Q^2) = \frac{F_{\tilde{D}}^2 M_{\tilde{D}}^4}
  {M_{\tilde{D}}^2 + Q^2},
\end{eqnarray}
where $F_D$ ($F_{\tilde{D}}$) denotes the $D$-meson
($\tilde{D}$-meson) decay constant and the space-like momentum
$Q^2 = (M_{D} + \Lambda_M)^2$ with $\Lambda_M$ being the matching scale.
We note that the heavy quark limit $M_D \to \infty$
should be taken with $\Lambda_M$ kept fixed
since $\Lambda_M$ must be smaller than the chiral symmetry breaking
scale
characterized by $\Lambda_\chi \sim 4\pi f_\pi$.
Then, $Q^2$ should be regarded as $Q^2 \simeq M_D^2$
in the present framework based on the chiral and heavy quark symmetries.
If we ignore the difference between $F_D$ and
$F_{\tilde{D}}$ which can be justified by the QCD sum rule
analysis~\cite{Narison:2003td}, then we get
 \bea
\Delta_{SP}(Q^2) \equiv G_S(Q^2) - G_P(Q^2)\simeq \frac{3 F_D^2
M_D^3}{M_D^2 + Q^2}
   \Delta M_D.
\label{diff-eft}
 \eea
In the QCD sector, the correlators $G_S$ and $G_P$ are given by
the operator product expansion (OPE) as~\cite{Narison:1988ep}
\begin{eqnarray}
  G_S(Q^2) &=&
  \left. G(Q^2) \right\vert_{\rm pert}\nonumber\\
   &+&
  \frac{m_H^2}{m_H^2 + Q^2}
  \Biggl[ {}- m_H \langle \bar{q}q \rangle +
   \frac{\alpha_s}{12\pi}\langle G^{\mu\nu}G_{\mu\nu} \rangle \Biggr],
\nonumber\\
  G_P(Q^2) &=&
  \left. G(Q^2) \right\vert_{\rm pert}\nonumber\\
   &+&
  \frac{m_H^2}{m_H^2 + Q^2}
  \Biggl[ m_H \langle \bar{q}q \rangle +
   \frac{\alpha_s}{12\pi}\langle G^{\mu\nu}G_{\mu\nu} \rangle \Biggr],
\end{eqnarray}
where $m_H$ is the heavy-quark mass. To the accuracy we are aiming
at, the OPE can be truncated at ${\mathcal O}(1/Q^2)$. The
explicit expression for the perturbative contribution $\left.
G(Q^2) \right\vert_{\rm pert}$ is available in the literature but
we do not need it since it drops out in the difference. {}From
these correlators, the $\Delta_{SP}$ becomes
\begin{equation}
 \Delta_{SP}(Q^2) = - \frac{2 m_H^3}{m_H^2 + Q^2}
  \langle \bar{q}q \rangle.
\label{diff-ope}
\end{equation}
Equating Eq.~(\ref{diff-eft}) to Eq.~(\ref{diff-ope}) and
neglecting the difference $(m_H-M_D)$, we obtain the following
matching condition:
\begin{equation}
 3 F_D^2 \, \Delta M \simeq - 2 \langle \bar{q}q \rangle.
\label{bare mass diff}
\end{equation}
Thus at the matching scale, the splitting is
\begin{equation}
 \Delta M_{\rm bare} \simeq
  -\frac{2}{3}\frac{\langle \bar{q}q \rangle}{F_D^2}.
 \label{bare diff}
\end{equation}
As announced, the $bare$ splitting is indeed proportional to the
light-quark condensate. The quantum corrections do not change the
dependence on the quark condensate [see section~\ref{sec:QC}].

%%%%%%%%%%%%%%%%%%%%%%%%%%%%%%%%%%%%%%%%%%%%%%%%%%%%%%%%%%%%%%%%%%%%%
%%%%%%%%%%%%%%%%%%%%%%%%%%%%%%%%%%%%%%%%%%%%%%%%%%%%%%%%%%%%%%%%%%%%%

\section{Quantum Corrections and RGE}
\label{sec:QC}

Given the bare Lagrangian whose parameters are fixed at the
matching scale $\Lambda_M$, the next step is to decimate the
theory \`a la Wilson to the scale at which $\Delta M$ is measured.
This amounts to calculating quantum corrections to the mass
difference $\Delta M$ in the framework of the present EFT.

\begin{figure}
\begin{center}
 \includegraphics
  [width = 6cm]
  {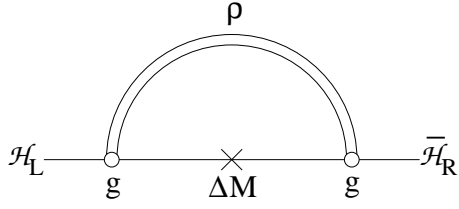}
\end{center}
\caption{Diagram contributing to the mass difference.}
\label{fig:mass}
\end{figure}

This calculation
turns out to be surprisingly simple for $a\approx 1$. If one sets
$a=1$ which is the approximation we are adopting here, $\alpha_L$
does not mix with $\alpha_R$ in the light sector, and then $\alpha_L$
couples to only $\calH_L$ and $\alpha_R$ to only $\calH_R$. As a
result $\calH_{L(R)}$ cannot connect to $\calH_{R(L)}$ by the
exchange of $\alpha_L$ or $\alpha_R$.
Only the $\rho$-loop links between the fields with different
chiralities as shown in Fig.~\ref{fig:mass}. We have verified this
approximation to be reliable since corrections to
the result with $a=1$ come only
at higher loop orders [see the next paragraph].
The diagram shown in Fig.~\ref{fig:mass} contributes to
the two-point function as
\begin{eqnarray}
 \biggl. \Pi_{LR} \biggr\vert_{\rm div}
 = - \frac{1}{2} \Delta M \,
   {\mathcal C}_2(N_f) \,
   \frac{g^2}{2\pi^2}\Bigl( 1 - 2k - k^2 \Bigr)
   \ln\Lambda,
\label{quantum correction}
\end{eqnarray}
where ${\mathcal C}_2(N_f)$ is the second Casimir defined by
$(T_a)_{ij}(T_a)_{jl}= {\mathcal C}_2(N_f)\delta_{il}$ with $i, j$
and $l$ denoting the flavor indices of the light quarks. This
divergence is renormalized by the bare contribution of the form
$\Pi_{LR,{\rm bare}} = \frac{1}{2}\Delta M_{\rm bare}$. Thus the
renormalization-group equation (RGE) takes the form
\begin{equation}
 \mu\frac{d\,\Delta M}{d\mu}
 =  {\mathcal C}_2(N_f)\,
   \frac{g^2}{2\pi^2}\Bigl( 1 - 2k - k^2 \Bigr)\Delta M.
\label{rge a=1}
\end{equation}
For an approximate estimate that we are interested in at this point,
it seems reasonable to ignore the scale dependence in $g$ and $k$.
Then the solution is simple:
\begin{equation}
 \Delta M = \Delta M_{{\rm bare}}
 \times C_{\rm quantum} \ ,
\label{mass diff}
\end{equation}
where we define $C_{\rm quantum}$ by
\begin{equation}
 C_{\rm quantum} =
 \exp\Bigl[ - {\mathcal C}_2(N_f)\,
   \frac{g^2}{2\pi^2}\Bigl( 1 - 2k - k^2 \Bigr)
   \ln\frac{\Lambda}{\mu} \Bigr]\ .
\label{Cquantum}
\end{equation}
This shows unequivocally that the mass
splitting is dictated by the ``bare" splitting $\Delta M_{\rm
bare}$ proportional to $\la\bar{q}q\ra$ corrected by the quantum
effect $C_{\rm quantum}$.

Next we lift the condition $a=1$ made in the above analysis. For
this purpose, we compute the quantum effects to the masses of
$0^-$ $(P)$ and $0^+$ $(Q^\ast)$ $D$-mesons by calculating the
one-loop corrections to the two-point functions of $P$ and
$Q^\ast$ denoted by $\Pi_{PP}$ and $\Pi_{Q^\ast Q^\ast}$ [for the
explicit calculation, see Appendix~\ref{app:ECQC}]. We find that
amazingly, the resultant form of the quantum correction exactly
agrees with the previous one which was obtained by taking $a=1$.
To arrive at this result, it is essential that $P$ (or
$P^\ast_\mu$) be the chiral partner of $Q^\ast$ (or $Q_\mu$) as
follows: The loop diagrams shown in Fig.~\ref{fig:PP} and
Fig.~\ref{fig:QQ} in Appendix~\ref{app:ECQC} have power and
logarithmic divergences. However all the divergences of the
diagrams with pion loop are exactly canceled among themselves
since the internal (or external) particles are chiral partners. In
a similar way, the exact cancellation takes place in the diagrams
with $\sigma$ loop. Finally, the logarithmic divergence from the
$\rho$ loop does contribute to the mass difference. This shows
that the effect of spontaneous chiral symmetry breaking introduced
as the deviation of $a$ from 1 does not get transferred to the
heavy sector. Thus even in the case of $a \neq 1$, the bare mass
splitting is enhanced by only the vector meson loop, with the
pions not figuring in the quantum corrections at least at one-loop
order. Solving the RGE (\ref{rge}), which is exactly same as
Eq.~(\ref{rge a=1}), we obtain exactly the same mass splitting as
the one given in Eq.~(\ref{mass diff}).

%%%%%%%%%%%%%%%%%%%%%%%%%%%%%%%%%%%%%%%%%%%%%%%%%%%%%%%%%%%%%%%%%%%%
%%%%%%%%%%%%%%%%%%%%%%%%%%%%%%%%%%%%%%%%%%%%%%%%%%%%%%%%%%%%%%%%%%%%

\section{Mass Splitting}
\label{sec:DMD}

In this section we make a numerical estimation of the mass
splitting for the chiral doublers in the open charm system. (Here
$D$ denotes the open charm meson.) Since we are considering the
chiral limit, strictly speaking, a precise comparison with
experiments is not feasible particularly if the light quark is
strange, so what we obtain should be considered as
semi-quantitative at best. This caveat should be kept in mind in
what follows.

Determining the bare mass splitting from the matching condition
(\ref{bare diff}) requires the quark condensate at that scale and
the $D$-meson decay constant $F_D$. For the quark condensate, we
shall use the so-called ``standard value"~\cite{GL}
$\langle\bar{q}q\rangle=-(225\pm 25\,\mbox{MeV})^3$ at $1$\,GeV.
Extrapolated to the scale $\Lambda_M=1.1$ GeV we shall adopt here,
this gives
 \begin{equation} \langle\bar{q}q\rangle_{\Lambda_M} =
-(228\pm25\,\mbox{MeV})^3.
 \end{equation}
Unfortunately this value is not firmly established, there being no
consensus on it. The values found in the literature vary widely,
even by a factor of $\sim 2$, some higher~\cite{BL} and some
lower~\cite{Moussallam}. (See Appendix \ref{quark condensate} for
more on this matter.) We shall therefore take the standard value
as a median~\footnote{
 It was shown in Ref.~\cite{Moussallam} that there is a
 strong $N_F$ dependence on the quark condensate and
 the value of the quark condensate for
 QCD with three massless quarks is smaller than
 the value used in estimating the value of the mass splitting
 in section~\ref{sec:DMD}.
 In the present analysis, we extract the value of the coupling
 constant $k$ from the experiment.  To be consistent, we need
 to use the quark condensate together with other parameters involved
 determined at the same scale from
 experimental and/or lattice data. This corresponds to the standard
value of the condensate we are using here.}.
As for the $D$-meson decay constant,  we take as a typical value
$F_D=0.205\pm 0.020\,\mbox{GeV}$ obtained from the QCD sum rule
analysis~\cite{Narison:2003td}. Plugging the above input values
into Eq.~(\ref{bare diff}) we obtain
\begin{equation}
\Delta M_{\rm bare} \simeq 0.19 \,\mbox{GeV} \ .
\end{equation}
By taking $\mu = m_\rho = 771\,\mbox{MeV}$,
$\Lambda = \Lambda_M = 1.1\,\mbox{GeV}$,
$g = g(m_\rho) = 6.27$ determined via the
Wilsonian matching for
$(\Lambda_M,\Lambda_{\rm QCD})=(1.1,0.4)\,\mbox{GeV}$
in Ref.~\cite{HY:PRep} and $k \simeq 0.59$
extracted from the $D^*\rightarrow D\pi$ decay
[see section~\ref{D*Dpi}] in Eq.~(\ref{Cquantum}),
we find for $N_F=3$
\begin{equation}
  C_{\rm quantum} = 1.6 \ .
\end{equation}
This is a sizable quantum correction involving only the vector
meson. If one takes into account the uncertainties involved in the
condensate and the decay constant, the quantum-corrected splitting
$\Delta M$ comes out to be
\begin{equation}
 \Delta M \approx 0.31\pm 0.12\,\mbox{GeV} \ .
\label{mass diff num}
\end{equation}
Despite the uncertainty involved, (\ref{mass diff num}) is a
pleasing result. It shows that the splitting is indeed of the size
of the constituent quark mass of a chiral quark $\Sigma \sim
m_p/3\sim 310$ MeV and is directly proportional to the quark
condensate.

We should stress however several caveats associated with this
result. Apart from the sensitivity to the quark condensate, if one
naively plugs in the matching scale $\Lambda_M$ into the RGE
solution, one finds the splitting is not insensitive as it should
be to the scale change. This is neither surprising nor too
disturbing since our RGE solution is obtained with the scale
dependence in both $g$ and $k$ ignored. In order to eliminate this
dependence on the matching scale, it will be necessary to solve
the RGE with the full scale dependence taken into account -- which
is at the moment beyond our scope here. The best we can do within
the scheme adopted is to pick the optimal $\Lambda_M$ determined
phenomenologically from elsewhere~\cite{HY:PRep} and this is what
we have done above.

%%%%%%%%%%%%%%%%%%%%%%%%%%%%%%%%%%%%%%%%%%%%%%%%%%%%%%%%%%%%%%%
%%%%%%%%%%%%%%%%%%%%%%%%%%%%%%%%%%%%%%%%%%%%%%%%%%%%%%%%%%%%%%%
\section{Hadronic Decay Modes}
\label{sec:HDM}

In this section we turn to the hadronic decay processes of the
$\tilde{D}$ mesons and make predictions of our scenario based on
the vector manifestation (VM) of chiral symmetry. Here we adopt
the notations $D_{u,d}$ and $\tilde{D}_{u,d}$ for the heavy
ground-state mesons and heavy excited mesons composed of
$c\bar{u}$ and $c\bar{d}$, and $D_{s}$ and $\tilde{D}_{s}$ for
those composed of $c\bar{s}$. The spin-parity quantum numbers will
be explicitly written as $D_{u,d}(0^-)$. For the heavy vector
meson, we follow the notation adopted by the Particle Data Group
(PDG)~\cite{Hagiwara:fs} and write $D_{u,d}^\ast(1^-)$ and
$D_{s}^\ast(1^-)$. Unless otherwise noted, the masses of the
ground-state heavy mesons will be denoted as $M_{D}$ and those of
the excited states as $M_{\tilde{D}}$.

\subsection{$D^\ast \rightarrow D + \pi$}
\label{D*Dpi}

Before studying the decay processes of the excited heavy mesons,
we first calculate the decay width of $D_{u,d}^\ast \rightarrow
D_{u,d} + \pi$ so as to determine the coupling constant $k$. The
decay widths of the $\pi^0$ and the $\pi^{\pm}$ modes are given by
\begin{eqnarray}
\Gamma(D_{u,d}^\ast(1^-) \rightarrow D_{u,d}(0^-) + \pi^0)
&=&
\frac{\bar{p}_\pi^3}{24\pi M_{D^\ast}^2 }
    \left( M_Q \, \frac{k}{F_\pi} \right)^2
\ ,
\nonumber\\
\Gamma(D^\ast_{u,d}(1^-) \rightarrow D_{u,d}(0^-) + \pi^{\pm})
&=&
\frac{\bar{p}_\pi^3}{12\pi M_{D^\ast}^2 }
    \left( M_Q \, \frac{k}{F_\pi} \right)^2
\ ,
\end{eqnarray}
where $\bar{p}_\pi \equiv |\vec{p}_\pi|$ denotes the
three-momentum of the
pion in the rest frame of the decaying particle $D^\ast_{u,d}(1^-)$,
and $M_Q$ the ``heavy quark mass'' introduced for
correctly normalizing the heavy meson field.
In the present analysis we use the following reduced mass
for definiteness:
\begin{equation}
 M_Q \equiv \frac{1}{4}\Bigl( M_{D(0^-)} + 3 M_{D^\ast(1^-)} \Bigr)
 = 1974 \, \mbox{MeV} \ .
\label{reduced mass}
\end{equation}

The total width is not determined for $D_u^\ast (1^-)$, although
the branching fractions for both the $\pi^0$ and the $\pi^+$ decay
modes are known experimentally. For $D_d^\ast (1^-)$ meson, on the
other hand, the total width is also determined.  Using the values
listed in PDG table~\cite{Hagiwara:fs}, the partial decay widths
are estimated to be
\begin{eqnarray}
&& \Gamma(D_{d}^\ast(1^-) \rightarrow D_{d}(0^-) + \pi^0)
  = 29.5 \pm 6.8 \,\mbox{keV}
\ ,
\nonumber\\
&& \Gamma(D_{d}^\ast(1^-) \rightarrow D_{u}(0^-) + \pi^+)
  = 65 \pm 15 \,\mbox{keV}
\ .
\end{eqnarray}
Here the $\pi^0$ mode will be used as an input to fix $k$. From
the experimental masses $M_{D_u^\ast (1^-)} = 2010.1$\,MeV,
$M_{D_d(0^-)} = 1869.4$\,MeV and $M_{\pi^0} = 134.9766$\,MeV
together with the value of the pion decay constant $F_\pi = 92.42
\pm 0.26$\,MeV, we obtain
\begin{equation}
k = 0.59 \pm 0.07 \ .
\end{equation}
Note that the error is mainly from that of the
$D_{d}^\ast(1^-) \rightarrow D_{d}(0^-) + \pi^0$ decay width.

In the following analysis, we shall use the central value of $k$
to make predictions for the decay widths of $\tilde{D}$ mesons.
Each prediction includes at least about 20\% error from the value
of $k$. For the masses of excited $D$ mesons, we use
$M_{\tilde{D_s}(0^+)}=2317$\,MeV determined by BaBar~\cite{BABAR},
$M_{\tilde{D_s}(1^+)}=2460$\,MeV by CLEO~\cite{CLEO} and
$(M_{\tilde{D}_{u,d}(0^+)},M_{\tilde{D}_{u,d}(1^+)})
=(2308,2427)$\,MeV by Belle~\cite{Belle}. Table~\ref{input}
summarizes the input parameters used in the present analysis.
\begin{table}
 \begin{center}
  \begin{tabular*}{16.8cm}{@{\extracolsep{\fill}}cccccc}
    \hline
    $D_{u,d}$ meson masses
     &  $M_{\tilde{D}_{u,d}(1^+)}$  &  $M_{\tilde{D}_{u,d}(0^+)}$
     &  $M_{D_{u,d}^\ast(1^-)}$     &  $M_{D_{u,d}(0^-)}$
     &  {}\\
    (MeV)
     &  2427                        &  2308
     &  2010                        &  1865
     &  {}\\
    \hline
    $D_s$ meson masses
     &  $M_{\tilde{D}_s(1^+)}$      &  $M_{\tilde{D}_s(0^+)}$
     &  $M_{D_s^\ast (1^-)}$        &  $M_{D_s(0^-)}$
     &  {}\\
    (MeV)
     &  2460                        &  2317
     &  2112                        &  1969
     &  {}\\
    \hline
    Light meson masses
     &  $M_\pi$          &  $M_\rho$         & $M_\eta$
     &  $M_\phi$          & {} \\
    (MeV)
     &  138.039          &  771.1            &  547.30
     &  1019.456         & {}\\
    \hline
    $\pi^0$-$\eta$ mixing
     &  $A_{11}$                         & $A_{21}$
     & $\Pi_{\pi^0\eta}\,(\mbox{MeV})^2$ &  $K_{\pi^0\eta}$       & {}\\
    {}
     &  0.71                             &  $-0.52$
     & $-4.25 \times 10^3$               & $-1.06 \times 10^{-2}$ & {}\\
    \hline
    $\phi$-$\rho$ mixing
     & $\Gamma_{\phi \to \pi^+\pi^-}\,$(MeV)
     & $\Gamma_{\rho \to \pi^+\pi^-}\,$(MeV)
     & {}                       & {}                     & {} \\
    {}
     & $3.11 \times 10^{-4}$  & 149.2
     & {}                     & {}                     & {}\\
    \hline
  \end{tabular*}
 \end{center}
 \caption{The values of input parameters.
  We use the values of $M_{\tilde{D}_s(0^+)}$~\cite{BABAR},
  $M_{\tilde{D}_s(1^+)}$~\cite{CLEO}
  and $M_{\tilde{D}_{u,d}(0^+,1^+)}$~\cite{Belle}.
  The $D$ mesons in the ground state, light mesons
  and decay widths $\Gamma (\phi, \rho)$ are
  the values listed by the PDG table~\cite{Hagiwara:fs}.
  As for the parameters associated with the $\pi^0$-$\eta$ mixing,
  we use the values given in
  Refs.~\cite{Schechter:1992iz,Harada:1995sj}.}
 \label{input}
\end{table}

\subsection{$\tilde{D} \to D + \pi$}

For the systems of $c\bar{u}$ and $c\bar{d}$,
the following decay processes of the $\tilde{D}_{u,d}$ meson
into the $D_{u,d}$ meson and one pion are allowed by the
spin and parity:
\begin{equation}
 \tilde{D}_{u,d}(0^+) \to D_{u,d}(0^-) + \pi
\qquad
 \tilde{D}_{u,d}(1^+) \to D_{u,d}^\ast (1^-) + \pi.
\end{equation}
Their partial decay widths are given by
\begin{eqnarray}
 && \Gamma (\tilde{D}_{u,d} \to D_{u,d} + \pi^{\pm})
   = \frac{\bar{p}_\pi}{4\pi}
   \Biggl( \frac{k}{F_\pi}\frac{M_Q}{M_{\tilde{D}}}E_\pi \Biggr)^2,
\nonumber\\
 && \Gamma (\tilde{D}_{u,d} \to D_{u,d} + \pi^0)
   = \frac{\bar{p}_\pi}{8\pi}
   \Biggl( \frac{k}{F_\pi}\frac{M_Q}{M_{\tilde{D}}}E_\pi \Biggr)^2,
\end{eqnarray}
where $E_\pi$ is the energy of the pion, and the reduced mass
$M_Q$ is defined in Eq.~(\ref{reduced mass}). With the input
parameters given in Table~\ref{input}, these decay widths come out
to be
\begin{eqnarray}
 && \Gamma (\tilde{D}_{u,d}(0^+) \to D_{u,d}(0^-) + \pi^0)
   = 73.7\,\mbox{MeV} \ ,
\nonumber\\
 && \Gamma (\tilde{D}_{u,d}(0^+) \to D_{u,d}(0^-) + \pi^{\pm})
   = 147 \,\mbox{MeV} \ ,
\nonumber\\
 && \Gamma (\tilde{D}_{u,d}(1^+) \to D_{u,d}^\ast (1^-) + \pi^0)
   = 57.2\,\mbox{MeV} \ ,
\nonumber\\
 && \Gamma (\tilde{D}_{u,d}(1^+) \to D_{u,d}^\ast (1^-) + \pi^{\pm})
   = 114 \,\mbox{MeV} \ ,
\end{eqnarray}

For the system of $c\bar{s}$ there are two decay processes of the
$\tilde{D}_s$ meson into the $D_s$ meson and one pion:
\begin{equation}
 \tilde{D}_s(0^+) \to D_s(0^-) + \pi^0
\qquad
 \tilde{D}_s(1^+) \to D_s^\ast (1^-) + \pi^0.
\end{equation}
These processes violate the isospin invariance, and hence are
suppressed. In the present analysis we assume as in
Ref.~\cite{Bardeen-Eichten-Hill:03} that the isospin violation
occurs dominantly through the $\pi^0$-$\eta$ mixing. In other
words, we assume that the $\tilde{D}_s$ meson decays into the
$D_s$ meson and the virtual $\eta$ meson which mixes with
the $\pi^0$
through the $\pi^0$-$\eta$ mixing.
Then, the decay width is given by
\begin{equation}
 \Gamma (\tilde{D}_s \to D_s + \pi^0)
 = \frac{\bar{p}_\pi}{2\pi}
   \Biggl( \frac{k}{F_\pi} \frac{M_Q}{M_{\tilde{D}}} E_\pi
     \Delta_{\pi^0\eta} \Biggr)^2,
\end{equation}
where $\Delta_{\pi^0 \eta}$ denotes the $\pi^0$-$\eta$ mixing
and takes the following form~\cite{Schechter:1992iz,Harada:1995sj}:
\begin{equation}
 \Delta_{\pi^0 \eta}
 = - \frac{A_{11}A_{21}}{M_\eta^2 - M_{\pi^0}^2}
   \bigl( \Pi_{\pi^0 \eta} - K_{\pi^0 \eta}M_{\pi^0}^2 \bigr)
\end{equation}
with $\Pi_{\pi^0 \eta}$ and $K_{\pi^0 \eta}$ being the mass-type and
kinetic-type $\pi^0$-$\eta$ mixing, respectively.
$A_{11}$ and $A_{21}$ are the components of the $\eta$-$\eta^\prime$
mixing matrix in the two-mixing-angle scheme~\cite{Schechter:1992iz}.
By using the values listed in Table~\ref{input},
the $\pi^0$-$\eta$ mixing is estimated as
\begin{equation}
 \Delta_{\pi^0 \eta} = -5.32 \times 10^{-3}.
\end{equation}
{}From this value, the decay widths are predicted as
\begin{eqnarray}
&& \Gamma(\tilde{D}_s(0^+) \to D_s(0^-) + \pi^0)
 = 4.17 \, \mbox{keV} \ ,
\nonumber\\
&& \Gamma(\tilde{D}_s(1^+) \to D_s^\ast (1^-) + \pi^0)
 = 3.75 \,\mbox{keV}
\ .
\end{eqnarray}

\subsection{$\tilde{D}(1^+) \rightarrow \tilde{D}(0^+) + \pi$}

With the masses of $\tilde{D}_{u,d}(1^+)$ and
$\tilde{D}_{u,d}(0^+)$ listed in Table~\ref{input}, the
intra-multiplet decay $\tilde{D}_{u,d}(1^+) \rightarrow
\tilde{D}_{u,d}(0^+) + \pi$ is not allowed kinematically. Since
the experimental errors for the masses are large~\footnote{
  The Belle collaboration~\cite{Abe:2003zm,Krokovny:2003rm,Belle:talk} gives
  $M_{\tilde{D}_{u,d}(1^+)} = 2427\pm26\pm20\pm17$\,MeV
  and $M_{\tilde{D}_{u,d}(0^+)} = 2308\pm17\pm15\pm28$\,MeV.
},
this decay mode
may still turn out to be possible. To show how large the possible
decay width is, we use $M_{\tilde{D}_{u,d}(0^+)} =
2272$\,\mbox{MeV} and $M_{\tilde{D}_{u,d}(1^+)} =
2464$\,\mbox{MeV} together with the formulas
\begin{eqnarray}
&&
\Gamma(\tilde{D}_{u,d}(1^+) \rightarrow \tilde{D}_{u,d}(0^+) + \pi^0 )
= \frac{\bar{p}_\pi^3}{24\pi}
    \left( \frac{M_Q}{M_{\tilde{D}(1^+)}} \,\frac{k}{F_\pi} \right)^2
\ ,
\nonumber\\
&&
\Gamma(\tilde{D}_{u,d}(1^+) \rightarrow \tilde{D}_{u,d}(0^+) + \pi^+ )
= \frac{\bar{p}_\pi^3}{12\pi}
    \left( \frac{M_Q}{M_{\tilde{D}(1^+)}} \,\frac{k}{F_\pi} \right)^2
\ .
\end{eqnarray}
The resultant decay widths are given by
\begin{eqnarray}
&&
\Gamma(\tilde{D}_{u,d}(1^+) \rightarrow \tilde{D}_{u,d}(0^+) + \pi^0 )
= 0.729\,\mbox{MeV}
\ ,
\nonumber\\
&&
\Gamma(\tilde{D}_{u,d}(1^+) \rightarrow \tilde{D}_{u,d}(0^+) + \pi^+ )
= 1.46\,\mbox{MeV}
\ .
\end{eqnarray}
They are smaller by the order of $10^{-2}$ than other one-pion modes
[see Table~\ref{predictions}].
This is caused by the suppression from the phase space.

With the present input values of $\tilde{D}$ masses, the process
$\tilde{D}_{s}(1^+) \rightarrow \tilde{D}_{s}(0^+) + \pi^0$ is
kinematically allowed. Similarly to the $\tilde{D}_s \to D_s +
\pi^0$ decay, we assume that this decay is dominated by the
process through the $\pi^0$-$\eta$ mixing. Then, the decay width
is given by
\begin{eqnarray}
 \Gamma (\tilde{D}_s(1^+) \to \tilde{D}_s(0^+) + \pi^0)
 &=& \frac{\bar{p}_\pi^3}{6\pi}
     \Biggl( \frac{k}{F_\pi}\frac{M_Q}{M_{\tilde{D}(1^+)}}
      \Delta_{\pi^0 \eta}
     \Biggr)^2
\nonumber\\
 &=& 1.87\times 10^{-3}\,\mbox{keV}.
\end{eqnarray}
This is very tiny due to the isospin violation
and the phase-space suppression.

\subsection{$\tilde{D} \to D + 2\pi$}

There are several processes such as $\tilde{D} \rightarrow D +
\pi^{\pm}\pi^{\mp}$ to which the light scalar mesons could give
important contributions. In models based on the standard scenario
of the chiral symmetry restoration in the light quark sector, the
scalar-meson coupling to the heavy-quark system is related to the
pion coupling, enabling one to compute the decay width. In our
model based on the VM of the chiral symmetry restoration, on the
other hand, it is the coupling constant of the vector meson to the
heavy system that is related to the pion coupling constant: Here
coupling of the scalar meson is not directly connected, at least
in the present framework which contains no explicit scalar
fields~\footnote{Scalar excitations can of course be generated at
high loop level to assure unitarity or with the account of QCD
trace anomaly but we shall not attempt this extension in this
paper.}, to do that of the pion. So, while we cannot make firm
predictions to processes for which scalar mesons might contribute,
we can make definite predictions on certain decay widths for which
scalar mesons do not figure. If one ignores isospin violation, the
two-pion decay processes $\tilde{D}_{u,d} \rightarrow D_{u,d} +
\pi^{\pm} \pi^0$ receive no contributions from scalar mesons. We
give predictions for these processes below. As for the two-pion
decay modes of the $\tilde{D}_s$ meson, the scalar mesons could
give a contribution. To have an idea, we shall also compute the
vector-meson contribution to this process.

First, consider $\tilde{D}_{u,d}(0^+) \to D_{u,d}^\ast (1^-) +
\pi^{\pm}\pi^0$. In this process, there are two contributions:
\begin{eqnarray}
 \tilde{D}_{u,d}(0^+) &\to& D_{u,d}^\ast (1^-) + \pi^{\pm}\pi^0
 \qquad\qquad\qquad\qquad\mbox{(direct)}
\nonumber\\
  &\,\,& D_{u,d}^\ast (1^-) + (\rho^{\pm} \to \pi^{\pm}\pi^0)
  \qquad\qquad\mbox{($\rho$-mediation)}
\ .
\end{eqnarray}
The decay width is given by
\begin{eqnarray}
 &&\Gamma (\tilde{D}_{u,d}(0^+) \to D_{u,d}^\ast (1^-) + \pi^\pm\pi^0)
\nonumber\\
 &&\quad =
     \frac{M_Q^2}{64(2\pi)^3 M_{\tilde{D}}^3}
     \frac{k^2}{F_\pi^4}
     \int dm_{D\pi}^2 \int dm_{\pi\pi}^2
     |F_{\tilde{D}D}|^2
\nonumber\\
 &&\qquad\times
     \Biggl[ m_{\pi\pi}^2 - 4M_\pi^2 + \frac{1}{4M_D^2}
      \bigl( m_{\pi\pi}^2 - M_{\tilde{D}}^2 - M_D^2 - 2M_\pi^2
       {}+ 2m_{D\pi}^2 \bigr)^2
     \Biggr],
\end{eqnarray}
with $m_{D\pi}^2 = (p_D + p_\pi)^2$ and $m_{\pi\pi}^2 = (p_{1\pi}
+ p_{2\pi})^2$. The form factor $F_{\tilde{D}D}$ is taken to be of
the form
\begin{eqnarray}
  F_{\tilde{D}D}
  = 1 + \frac{M_\rho^2}{m_{\pi\pi}^2 - M_\rho^2}
  \ .
\end{eqnarray}
The first term of the form factor comes from the direct
contribution and the second from the $\rho$-mediation. Here we
have neglected the $\rho$ meson width in the propagator, since the
maximum value of $m_{\pi\pi}$ is about $300$\,MeV with the input
values listed in Table~\ref{input}. We can see that the form
factor $F_{\tilde{D}D}$ vanishes in the limit of $m_{\pi\pi} \to
0$, which is a consequence of chiral symmetry~\footnote{
  It should be stressed that this cancellation occurs
  because the vector meson is included consistently with
  chiral symmetry, and that it is $not$ a
  specific feature of the VM.  The chiral symmetry
  restoration based on the VM implies that the coupling
  constant of the vector meson to the heavy system is
  equal to that of the pion.
}. We note that $\left. m_{\pi\pi} \right\vert_{\rm max} \simeq
300$\,MeV makes this decay width strongly suppressed due to the
large cancellation between the direct and $\rho$-mediated
contributions. Furthermore, since $300$\,MeV is close to the
two-pion threshold, additional suppression comes from the phase
space. Due to these two types of suppressions the predicted decay
width is predicted to be very small, of the order of $10^{-2}$
keV.~\footnote{
  Note that the prediction on the decay width is very sensitive
  to the precise value of the mass of $\tilde{D}(0^+)$ meson.
}

Next we consider the process $\tilde{D}_{u,d}(1^+) \to
D_{u,d}^\ast (1^-) + \pi\pi$. Again there are two contributions,
direct and a $\rho$-mediated:
\begin{eqnarray}
 \tilde{D}_{u,d}(1^+) &\to& D_{u,d}^\ast (1^-) + \pi^{\pm}\pi^0
  \qquad\qquad\qquad\qquad\mbox{(direct)}
\nonumber\\
  &\,\,& D_{u,d}^\ast (1^-) + (\rho^{\pm} \to \pi^{\pm}\pi^0)
  \qquad\qquad\mbox{($\rho$-mediation)}
\end{eqnarray}
The resultant decay width is given by
\begin{eqnarray}
 &&\Gamma (\tilde{D}_{u,d}(1^+) \to D_{u,d}^\ast (1^-) + \pi^{\pm}\pi^0)
\nonumber\\
 &&\quad =
     \frac{M_Q^2}{96(2\pi)^3 M_{\tilde{D}}^3}
     \frac{k^2}{F_\pi^4}
     \int dm_{D\pi}^2 \int dm_{\pi\pi}^2
     |F_{\tilde{D}D}|^2
\nonumber\\
 &&\qquad\times
     \Biggl[ m_{\pi\pi}^2 - 4M_\pi^2 + \frac{1}{4M_{\tilde{D}}^2}
      \bigl( m_{\pi\pi}^2 - M_{\tilde{D}}^2 - M_D^2 - 2M_\pi^2
       {}+ 2m_{D\pi}^2 \bigr)^2
     \Biggr].
\end{eqnarray}
Similarly to $\Gamma (\tilde{D}_{u,d}(0^+) \to D_{u,d}^\ast
(1^-)+\pi^{\pm}\pi^0)$, the width is again suppressed due to the
large cancellation between the direct and $\rho$-mediated
contributions. The suppression from the phase space, on the other
hand, is not so large
since $\left. m_{\pi\pi} \right\vert_{\rm max} \simeq 420$\,MeV is
not so close to the two-pion threshold. The resulting decay width
is
\begin{equation}
\Gamma (\tilde{D}_{u,d}(1^+) \to D_{u,d}^\ast (1^-) + \pi^{\pm}\pi^0)
= 11.8 \, \mbox{keV} \ .
\end{equation}

The decay width of the process
\begin{eqnarray}
 \tilde{D}_{u,d}(1^+) &\to& D_{u,d}(0^-) + \pi^{\pm}\pi^0
  \qquad\qquad\qquad\qquad\mbox{(direct)}
\nonumber\\
  &\,\,& D_{u,d}(0^-) + (\rho^{\pm} \to \pi^{\pm}\pi^0)
  \qquad\qquad\mbox{($\rho$-mediation)}
\end{eqnarray}
is given by
\begin{eqnarray}
 &&\Gamma (\tilde{D}_{u,d}(1^+) \to D_{u,d}(0^-) + \pi^\pm \pi^0)
\nonumber\\
 &&\quad =
     \frac{M_Q^2}{192(2\pi)^3 M_{\tilde{D}}^3}
     \frac{k^2}{F_\pi^4}
     \int dm_{D\pi}^2 \int dm_{\pi\pi}^2
     |F_{\tilde{D}D}|^2
\nonumber\\
 &&\qquad\times
     \Biggl[ m_{\pi\pi}^2 - 4M_\pi^2 + \frac{1}{4M_{\tilde{D}}^2}
      \bigl( m_{\pi\pi}^2 - M_{\tilde{D}}^2 - M_D^2 - 2M_\pi^2
       {}+ 2m_{D\pi}^2 \bigr)^2
     \Biggr].
\end{eqnarray}
In the present case, $m_{\pi\pi}|_{\rm max}\simeq 560\,\mbox{MeV}$
is much larger than the two-pion threshold and hence the width
becomes larger than other two-pion processes. We find
\begin{equation}
 \Gamma (\tilde{D}_{u,d}(1^+) \to D_{u,d}(0^-) + \pi^\pm \pi^0)
   = 314\,\mbox{keV}.
\end{equation}

Finally we turn to the decay $\tilde{D}_s(1^+) \to D_s(0^-) +
\pi^+\pi^-$ which as mentioned could receive direct contributions
from scalar excitations. Since we have not incorporated scalar
degrees of freedom in the theory, we might not be able to make a
reliable estimate even if were to go to higher-loop orders. Just
to have an idea as to how important the vector meson contribution
can be, we calculate the decay width in which the $\tilde{D_s}$
meson decays into two pions through the $\phi$ meson. This isospin
violating decay can occur through the direct $\phi$-$\pi$-$\pi$
coupling and the $\phi$-$\rho$ mixing:
\begin{eqnarray}
 \tilde{D}_s(1^+) &\to& D_s(0^-) + (\phi \to \pi^+\pi^-)
  \qquad\qquad\qquad\quad\mbox{(direct)}
\nonumber\\
  &\,\,& D_s(0^-) + (\phi \to \rho^0 \to \pi^+\pi^-)
  \qquad\qquad\mbox{($\phi$-$\rho$ mixing)}
\end{eqnarray}
Since the main contribution to the $\phi \rightarrow \pi\pi$ is
expected to be given by the $\phi$-$\rho$ mixing, we shall neglect
the direct $\phi$-$\pi$-$\pi$-coupling contribution in the
following. Then the decay width is given by
\begin{eqnarray}
 &&\Gamma (\tilde{D}_s(1^+) \to D_s(0^-) + \pi^+\pi^-)
\nonumber\\
 &&\quad =
     \frac{M_Q^2}{192(2\pi)^3 M_{\tilde{D}}^3}
     \frac{k^2}{F_\pi^4}
     \int dm_{D\pi}^2 \int dm_{\pi\pi}^2
     \Biggl[ \frac{M_\rho^2 \Pi_{\phi\rho}}
      {(m_{\pi\pi}^2 - M_\phi^2)(m_{\pi\pi}^2 - M_\rho^2)}
     \Biggr]^2
\nonumber\\
 &&\qquad\times
     \Biggl[ m_{\pi\pi}^2 - 4M_\pi^2 + \frac{1}{4M_{\tilde{D}}^2}
      \bigl( m_{\pi\pi}^2 - M_{\tilde{D}}^2 - M_D^2 - 2M_\pi^2
       {}+ 2m_{D\pi}^2 \bigr)^2
     \Biggr],
\end{eqnarray}
where $\Pi_{\phi\rho}$ denotes the $\phi$-$\rho$ mixing
given by
\begin{equation}
 \Pi_{\phi\rho}^2
 = (M_\phi^2 - M_\rho^2)^2
   \Biggl( \frac{\bar{p}_\pi (\rho)}{\bar{p}_\pi (\phi)} \Biggr)^3
   \frac{M_\phi^2}{M_\rho^2}
   \frac{\Gamma (\phi \to \pi^+\pi^-)}{\Gamma (\rho \to \pi^+\pi^-)},
\end{equation}
with $\bar{p}_\pi (X)$ being the three-momentum of pion in the
rest frame of the decaying particle $X = \phi, \rho$. Using the
values listed in Table~\ref{input}, we have
\begin{equation}
\Pi_{\phi\rho} = 530\,(\mbox{MeV})^2
\end{equation}
so the decay width is predicted to be
\begin{equation}
\Gamma (\tilde{D}_s(1^+) \to D_s(0^-) + \pi^+\pi^-)
 = 2.13 \times 10^{-4} \,\mbox{keV} \ .
\end{equation}
The $\phi$-$\rho$ mixing is caused by the isospin violation, and
this process is highly suppressed. We conclude that should a
measured width come out to be substantially greater than what we
found here, it would mean that either scalars must figure
importantly or the VM is invalid in its present form.

\subsection{Summary of hadronic decay modes}

Our predictions of the decay widths are summarized in
Table~\ref{predictions}.
\begin{table}
 \begin{center}
  \begin{tabular*}{15cm}{@{\extracolsep{\fill}}cll}
    \hline
    Decaying particle       & Process
    & Width (MeV) \\
    \hline
    $\tilde{D}_{u,d}$       & $0^+ \to 0^- + \pi^0$
     & $7.37\times 10^1$\\
    {}                      & $0^+ \to 0^- + \pi^\pm$
     & $1.47\times 10^2$\\
    {}                      & $0^+ \to 1^- + \pi^\pm\pi^0$
     & $1.54\times 10^{-5}$\\
    {}                      & $1^+ \to 1^- + \pi^0$
     & $5.72\times 10^1$\\
    {}                      & $1^+ \to 1^- + \pi^\pm$
     & $1.14\times 10^2$\\
    {}                      & $1^+ \to 0^- + \pi^\pm\pi^0$
     & $3.14\times 10^{-1}$\\
    {}                      & $1^+ \to 1^- + \pi^\pm\pi^0$
     & $1.18\times 10^{-2}$\\
    \hline
    $\tilde{D}_s$           & $0^+ \to 0^- + \pi^0$
     & $4.17\times 10^{-3}$\\
    {}                      & $1^+ \to 0^+ + \pi^0$
     & $1.87\times 10^{-6}$\\
    {}                      & $1^+ \to 1^- + \pi^0$
     & $3.75\times 10^{-3}$\\
    {}                      & $1^+ \to 0^- + \pi^+\pi^-
       \quad (\mbox{through}\,\phi \to \rho^0 \to \pi^+\pi^-)$
     & $2.13\times 10^{-7}$\\
    \hline
  \end{tabular*}
 \end{center}
 \caption{The predicted values of the hadronic decay processes.}
 \label{predictions}
\end{table}
It should be stressed that the values obtained in this paper on
the one-pion reflect only that the $\tilde{D}$ meson is the chiral
partner of the $D$ meson. They are $not$ specific to the VM. We
therefore expect that as far as the one-pion processes are
concerned, there will be no essential differences between our
predictions and those in Ref.~\cite{Bardeen-Eichten-Hill:03}.
However, in the two-pion decay processes in which the scalar meson
does not mediate, our scenario based on the VM can make definite
predictions which might be distinguished from that based on the
standard picture. Especially for $\tilde{D}_{u,d}(1^+) \to
D_{u,d}(0^-) + \pi^{\pm}\pi^0$, we obtain a larger width than for
other two-pion modes. Although the predicted width is still small
-- perhaps too small to be detected experimentally, it is
important because of the following reason. In our approach, since
the excited heavy meson multiplets of $\tilde{D}(0^+)$ and
$\tilde{D}(1^+)$ denoted by $G$ are the chiral partners to the
ground-state multiplets denote by $H$, the $G$-$\bar{H}$-$\pi$
coupling is the same as the $H$-$\bar{H}$-$\pi$ coupling [see the
fifth and first terms of Eq.~(\ref{Lag:component})]. Thus the
width which is dependent on the strength of $k$ is a good probe to
test our scenario. The common $k$ is also essential for the ratio
of the widths of the two-pion modes to those of the one-pion
modes, which has no $k$ dependence. These are therefore are
definite predictions of our scenario. {}From the values listed in
Table~\ref{predictions}, we obtain
\begin{equation}
 \frac{\Gamma (\tilde{D}_{u,d}(1^+) \to D_{u,d}(0^-) + \pi^\pm\pi^0)}
      {\Gamma_{\pi + 2\pi}^{\rm (had)}}
 = 1.83 \times 10^{-3},
\end{equation}
where $\Gamma_{\pi + 2\pi}^{\rm (had)}$ is the sum of the widths
of the one-pion and two-pion modes of the decaying
$\tilde{D}_{u,d}(1^+)$.

%%%%%%%%%%%%%%%%%%%%%%%%%%%%%%%%%%%%%%%%%%%%%%%%%%%%%%%%%%%%%%%%%%%%
%%%%%%%%%%%%%%%%%%%%%%%%%%%%%%%%%%%%%%%%%%%%%%%%%%%%%%%%%%%%%%%%%%%%

\section{Summary and Discussions}
\label{sec:SD}

Let us summarize what we have accomplished in this paper. In
Ref.~\cite{Nowak-Rho-Zahed:03}, it was suggested that the chiral
doubling of the heavy-light mesons could be exploited as a litmus
indicator for chiral symmetry restoration by measuring the
splitting at high temperature or density. If the splitting is
indeed tied to the quark condensate which is an order parameter of
chiral symmetry, one could observe the splitting disappearing at
the critical point $C_\chi$. In this paper, we go the other way
around. We start by the observation that at the critical point,
the vector manifestation (VM) is realized~\cite{HYb,HY:PRep} so
that hadron masses vanish in a manner predicted by BR
scaling~\cite{BR91}. By introducing the deviation from the VM
fixed point in terms of chiral symmetry breaking in the
heavy-light system and matching the EFT so constructed to QCD at
the matching scale, the ``bare" mass splitting of the chiral
doublers is determined in terms of the quark condensate and other
QCD parameters of the system. The physical splitting is then
determined by doing renormalization group evolution of the
parameters with the bare Lagrangian matched to QCD. It is found to
reproduce semi-quantitatively -- modulo the spin assignments --
the observed splitting which is related to the constituent quark
mass. This result suggests rather strongly that identifying the
chiral restoration as the VM fixed point and the chiral doubling
as a signal of spontaneous breaking of chiral symmetry are
mutually consistent.

One of the significant results of the analysis presented in this
paper is that the vector meson plays an important role in
accounting for the splitting in the $D$ and $\tilde{D}$ mesons:
The {\it bare} mass splitting determined through the matching is
estimated as about $190$ MeV, too small to explain the observed
mass difference. However by including the quantum corrections
through the hadronic loop, the bare mass splitting is enhanced by
$\sim60\%$, where only the loop effect of the vector meson
contributes to the running of the mass splitting. The
contributions from the pion loop are completely cancelled among
themselves. This implies that the observed mass difference can not
be understood if one takes only the pion as the relevant degree of
freedom and that we need other degrees of freedom. In the VM, it
is nothing but the vector meson. The situation here is much like
in the calculation of pion velocity at the chiral restoration
point: The pion velocity is zero if the pion is the only effective
degree of freedom but approaches 1 if the vector meson with the VM
is included~\cite{HKRS}.

Moreover, the result is independent of the deviation of $a$ from
the fixed point value 1 at one-loop level. In other word, the
resultant form of the quantum correction at one-loop level is
completely independent of $a$. This implies that the deviation of
$a$ from 1 which reflects the effect of spontaneous chiral
symmetry breaking in the light quark sector does not get
transmitted to the heavy sector. This strongly suggests that the
deviation from $a=1$ involves physics that is not as primary as
the non-vanishing gauge coupling $g \neq 0$ in the description of
the broken phase: The deviation seems to be a ``secondary''
phenomenon, which is generated from $g \neq 0$ as expected in
Refs.~\cite{Georgi:gp, Georgi:1989xy}.~\footnote
{ Although $a=1$
  is the fixed point of the RGE at one-loop level, the deviation of
  $a$ from $1$ is generated by the finite renormalization part
  once we allow the deviation of the gauge coupling $g$
  from $0$~\cite{VVD:HS}. }
In fact, even
when we start from the HLS theory with $g \neq 0$ and $a=1$, the
physical quantities obtained through the Wilsonian matching are in
good agreement with experimental results as discussed in
~\cite{HY:PRep}. This observation supports the above argument. It
is intriguing to note that $a\sim 1$ is realized in the structure
of both non-exotic and exotic baryons such as the nucleon
electromagnetic form factor~\cite{DD} and the skyrmion description
of the $\Theta^+$ pentaquark~\cite{prm-penta}.

In section~\ref{sec:HDM}
we studied the hadronic decay processes
of the $\tilde{D}$
mesons and showed the predictions of our scenario.
The predictions
on the one-pion processes
are the consequences
of the fact that the $\tilde{D}$ meson is the chiral partner
to the $D$ meson, and
there are no essential differences between our predictions
and those in Ref.~\cite{Bardeen-Eichten-Hill:03}.
On the other hand,
in the two-pion decay processes in which the scalar meson
does not mediate, our scenario gives definite predictions,
since the vector meson coupling to the heavy system is
equivalent to the pion coupling due to the VM.
Although the predicted values of widths are small,
we hope that they are clarified in future experiment.

Several comments are in order:

In this paper, we introduced spontaneous chiral symmetry breaking
in the heavy sector by $\Delta M_{\rm bare}$ only.
Although
we took the common coefficient $k$ for all the interaction terms
in Eq.~(\ref{Lag:component}),
each interaction term generally has its own
coefficient different from others. However, we expect that
the effect of these interaction terms is suppressed by the factor
$1/\Lambda$ and as a result the contribution to $\Delta M$ is small
since the dimension of them is higher than that of the mass term.

It is interesting that the bare splitting depends on the
heavy-meson decay constant. This suggests that the splitting may
show heavy-quark flavor dependence. This could be checked with
experiments once a systematic heavy-quark expansion (which is not
done here) is carried out. It is only in this sense that (part of)
the splitting can be identified with the light-quark constituent
mass discussed in Ref.~\cite{Nowak-Rho-Zahed:93,Bardeen-Hill:94}.

%%%%%%%%%%%%%%%%%%%%%%%%%%%%%%%%%%%%%%%%%%%%%%%%%%%%%%%%%%%%%%%%%%%
%%%%%%%%%%%%%%%%%%%%%%%%%%%%%%%%%%%%%%%%%%%%%%%%%%%%%%%%%%%%%%%%%%%

\section*{Acknowledgment}

One of the authors (MR) is grateful for illuminating discussions
with Maciek Nowak.
MH and CS would like to thank the members of Korea Institute for
Advanced Study (KIAS) for the warm hospitality
where this work was partially done.
The work of MH and CS are supported in part by the 21st Century COE
Program of Nagoya University provided by Japan Society for the
Promotion of Science (15COEG01), and by the JSPS Grant-in-Aid for
Scientific Research (c) (2) 16540241.

%%%%%%%%%%%%%%%%%%%%%%%%%%%%%%%%%%%%%%%%%%%%%%%%%%%%%%%%%%%%%%%%%%%%
%%%%%%%%%%%%%%%%%%%%%%%%%%%%%%%%%%%%%%%%%%%%%%%%%%%%%%%%%%%%%%%%%%%%

\appendix

\setcounter{section}{0}
\renewcommand{\thesection}{\Alph{section}}
\setcounter{equation}{0}
\renewcommand{\theequation}{\Alph{section}.\arabic{equation}}

\section{Explicit Calculation of Quantum Correction}
\label{app:ECQC}

In this appendix, we compute the quantum effects on the masses of
$0^-$ $(P)$ and $0^+$ $(Q^\ast)$ heavy-light $\calM$-mesons by
calculating the one-loop corrections to the two-point functions of
$P$ and $Q^\ast$ denoted by $\Pi_{PP}$ and $\Pi_{Q^\ast Q^\ast}$.
Here we adopt the following regularization method to identify the
power divergences: We first perform the integration over the
temporal component of the integration momentum, and then in the
remaining integration over three-momentum we make the replacements
given by
\begin{equation}
\int^\Lambda \frac{d^3\vec{k}}{(2\pi)^3} \frac{1}{\bar{k}^2}
\ \rightarrow \ \frac{\Lambda}{2\sqrt{2} \pi^2}
\ ,
\quad
\int^\Lambda \frac{d^3\vec{k}}{(2\pi)^3} \frac{1}{\bar{k}}
\ \rightarrow \ \frac{\Lambda^2}{8 \pi^2}
\ ,
\quad
\int^\Lambda \frac{d^3\vec{k}}{(2\pi)^3}
\ \rightarrow\  \frac{\Lambda^3}{12\sqrt{2}\pi^2}
\ .
\label{p divs}
\end{equation}
Here we use the t'Hooft-Feynman gauge for fixing the gauge of the
HLS.

\begin{figure}
\begin{center}
 \includegraphics
  [width = 13cm]
  {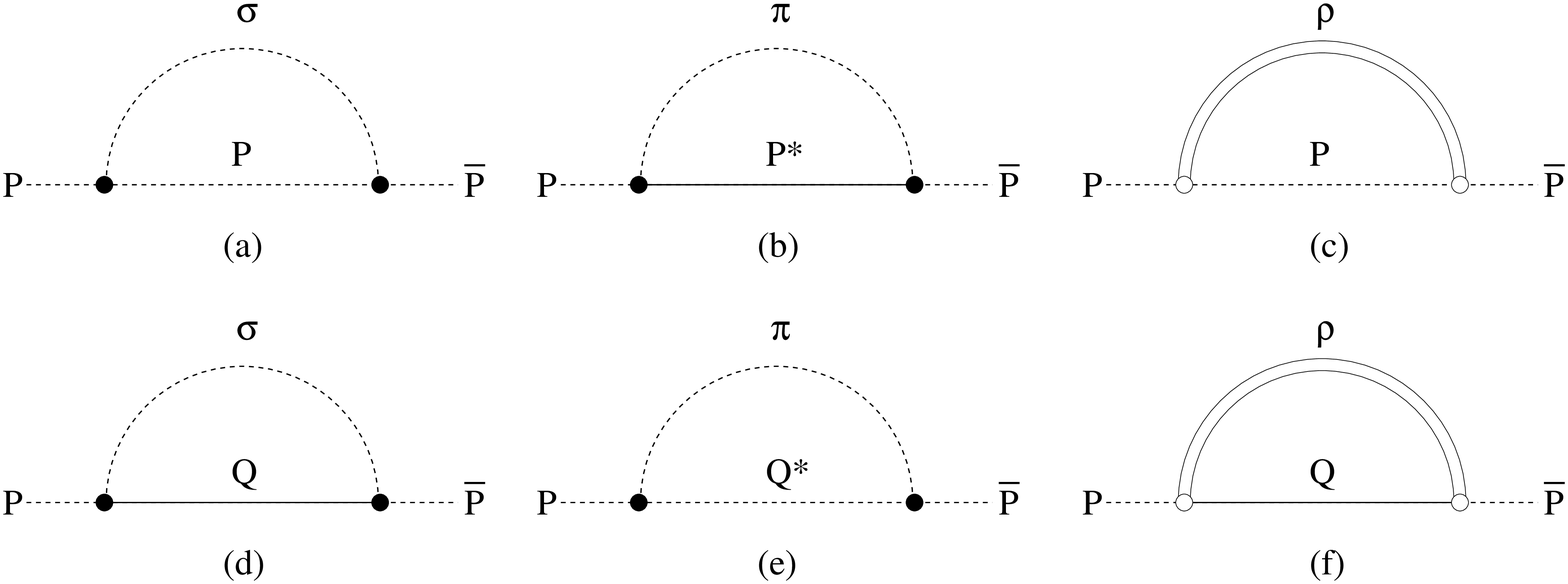}
\end{center}
\caption{Diagrams contributing to $P$-$P$ two point function.}
\label{fig:PP}
\end{figure}

The diagrams contributing to $\Pi_{PP}$ are shown in
Fig.~\ref{fig:PP}.
In the limit of zero external momentum, the divergent parts of
these contributions are given by
\begin{eqnarray}
&& \biggl. \Pi_{PP}^{(a)[\sigma P]} \biggr\vert_{\rm div}
  = \frac{2k^2}{F_\sigma^2}
  \left[
    - \frac{M_H}{(4\pi)^2}
      \left( \Lambda^2 - 2 M_\rho \ln \Lambda \right)
    + \frac{M_H^2}{4\pi^2}
      \left( \frac{\Lambda}{\sqrt{2}} - M_H \ln \Lambda \right)
  \right]
\ ,
\nonumber\\
&& \biggl. \Pi_{PP}^{(b)[\pi P^\ast]} \biggr\vert_{\rm div}
  = \frac{2k^2}{F_\pi^2}
  \left[
    \frac{\Lambda^3}{24\sqrt{2}\pi^2}
    - \frac{M_H}{(4\pi)^2} \Lambda^2
    + \frac{M_H^2}{4\pi^2}
      \left( \frac{\Lambda}{\sqrt{2}} - M_H \ln \Lambda \right)
  \right]
\ ,
\nonumber\\
&& \biggl. \Pi_{PP}^{(c)[\rho P]} \biggr\vert_{\rm div}
  = \frac{g^2}{2\pi^2}  \left( 1 - k \right)^2
  \left( \frac{\Lambda}{\sqrt{2}} - M_H \ln \Lambda \right)
\ ,
\nonumber\\
&& \biggl. \Pi_{PP}^{(d)[\sigma Q]} \biggr\vert_{\rm div}
  = \frac{2k^2}{F_\sigma^2}
  \left[
    \frac{\Lambda^3}{24\sqrt{2}\pi^2}
    - \frac{M_G}{(4\pi)^2}
      \left( \Lambda^2 - 2 M_\rho^2 \ln \Lambda \right)
    + \frac{M_G^2-M_\rho^2}{4\pi^2}
      \left( \frac{\Lambda}{\sqrt{2}} - M_G \ln \Lambda \right)
  \right]
\ ,
\nonumber\\
&& \biggl. \Pi_{PP}^{(e)[\pi Q^\ast]} \biggr\vert_{\rm div}
  = \frac{2k^2}{F_\pi^2}
  \left[
    - \frac{M_G}{(4\pi)^2} \Lambda^2
    + \frac{M_G^2}{4\pi^2}
      \left( \frac{\Lambda}{\sqrt{2}} - M_G \ln \Lambda \right)
  \right]
\ ,
\nonumber\\
&& \biggl. \Pi_{PP}^{(f)[\rho Q]} \biggr\vert_{\rm div}
  = \frac{3 g^2}{2\pi^2} k^2
  \left( \frac{\Lambda}{\sqrt{2}} - M_G \ln \Lambda \right)
\ . \label{Pi PP}
\end{eqnarray}
The particles that figure in the loop are indicated by the suffix
in square bracket; e.g., $[\pi P^\ast]$ indicates that $\pi$ and
$P^\ast$ enter in the internal lines. Here and henceforth, we
suppress, for notational simplification, the group factor
${\mathcal C}_2(N_f)$ defined as $(T_a)_{ij}(T_a)_{jl} = {\mathcal
C}_2(N_f)\delta_{il}$.

\begin{figure}
\begin{center}
 \includegraphics
  [width = 13cm]
  {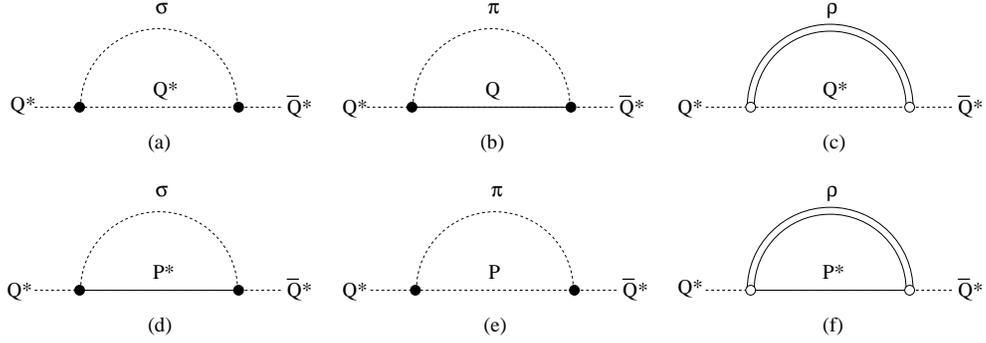}
\end{center}
\caption{Diagrams contributing to $Q^\ast$-$Q^\ast$ two point
function.}
\label{fig:QQ}
\end{figure}

The relevant diagrams contributing to
$\Pi_{Q^\ast Q^\ast}$ are
shown in Fig.~\ref{fig:QQ}.
The divergent parts of
these contributions in the low-energy limit are expressed as
\begin{eqnarray}
&& \biggl. \Pi_{Q^\ast Q^\ast}^{(a)[\sigma Q^\ast]}
  \biggr\vert_{\rm div}
  = \frac{2k^2}{F_\sigma^2}
  \left[
    - \frac{M_G}{(4\pi)^2}
      \left( \Lambda^2 - 2 M_\rho^2 \ln \Lambda \right)
    + \frac{M_G^2}{4\pi^2}
      \left( \frac{\Lambda}{\sqrt{2}} - M_G \ln \Lambda \right)
  \right]
\ ,
\nonumber\\
&& \biggl. \Pi_{Q^\ast Q^\ast}^{(b)[\pi Q]} \biggr\vert_{\rm div}
  = \frac{2k^2}{F_\pi^2}
  \left[
    \frac{\Lambda^3}{24\sqrt{2}\pi^2}
    - \frac{M_G}{(4\pi)^2} \Lambda^2
    + \frac{M_G^2}{4\pi^2}
      \left( \frac{\Lambda}{\sqrt{2}} - M_G \ln \Lambda \right)
  \right]
\ ,
\nonumber\\
&& \biggl. \Pi_{Q^\ast Q^\ast}^{(c)[\rho Q^\ast]}
  \biggr\vert_{\rm div}
  = \frac{g^2}{2\pi^2}  \left( 1 - k \right)^2
  \left( \frac{\Lambda}{\sqrt{2}} - M_G \ln \Lambda \right)
\ ,
\nonumber\\
&& \biggl. \Pi_{Q^\ast Q^\ast}^{(d)[\sigma P^\ast]}
  \biggr\vert_{\rm div}
  = \frac{2k^2}{F_\sigma^2}
  \left[
    \frac{\Lambda^3}{24\sqrt{2}\pi^2}
    - \frac{M_H}{(4\pi)^2}
      \left( \Lambda^2 - 2 M_\rho^2 \ln \Lambda \right)
    + \frac{M_H^2-M_\rho^2}{4\pi^2}
      \left( \frac{\Lambda}{\sqrt{2}} - M_H \ln \Lambda \right)
  \right]
\ ,
\nonumber\\
&& \biggl. \Pi_{Q^\ast Q^\ast}^{(e)[\pi P]} \biggr\vert_{\rm div}
  = \frac{2k^2}{F_\pi^2}
  \left[
    - \frac{M_H}{(4\pi)^2} \Lambda^2
    + \frac{M_H^2}{4\pi^2}
      \left( \frac{\Lambda}{\sqrt{2}} - M_H \ln \Lambda \right)
  \right]
\ ,
\nonumber\\
&& \biggl. \Pi_{Q^\ast Q^\ast}^{(f)[\rho P^\ast]} \biggr\vert_{\rm div}
  = \frac{3 g^2}{2\pi^2} k^2
  \left( \frac{\Lambda}{\sqrt{2}} - M_H \ln \Lambda \right)
\ .
\label{Pi QQ}
\end{eqnarray}

Now, let us compute the difference of $\Pi_{Q^\ast Q^\ast} - \Pi_{PP}$.

It is easy to show that $\left. \Pi_{PP}^{(b+e)}\right\vert_{\rm
div}$ exactly cancels with $\left. \Pi_{Q^\ast Q^\ast}^{(b+e)}
\right\vert_{\rm div}$. {}From the explicit forms given in
Eqs.~(\ref{Pi PP}) and (\ref{Pi QQ}), we have
\begin{eqnarray}
&& \biggl. \Pi_{Q^\ast Q^\ast}^{(b)[\pi Q]}
   - \Pi_{PP}^{(e)[\pi Q^\ast]} \biggr\vert_{\rm div}
  = \frac{2k^2}{F_\pi^2} \frac{\Lambda^3}{24\sqrt{2}\pi^2}
\ ,
\nonumber\\
&& \biggl. \Pi_{Q^\ast Q^\ast}^{(e)[\pi P]}
   - \Pi_{PP}^{(b)[\pi P^\ast]} \biggr\vert_{\rm div}
  = - \frac{2k^2}{F_\pi^2} \frac{\Lambda^3}{24\sqrt{2}\pi^2}
\ .
\end{eqnarray}
Note that the logarithmic, linear and quadratic divergences in
$\Pi_{Q^\ast Q^\ast}$ are exactly canceled by those in $\Pi_{PP}$.
This cancellation simply reflects that the external particles are
chiral partners. This immediately leads to
\begin{equation}
\biggl. \Pi_{Q^\ast Q^\ast}^{(b+e)} - \Pi_{PP}^{(b+e)}\biggr\vert_{\rm div} = 0
\ .
\label{QQPP be}
\end{equation}
The cubic divergence in $\Pi_{Q^\ast Q^\ast}$ is exactly canceled
by that in $\Pi_{PP}$, reflecting that the internal particles are
chiral partners to each other.

In a similar way, a partial cancellation takes place between
$\Pi_{Q^\ast Q^\ast}^{(a)}$ and $\Pi_{PP}^{(d)}$ as well as
between $\Pi_{Q^\ast Q^\ast}^{(d)}$ and $\Pi_{PP}^{(a)}$:
\begin{eqnarray}
&& \biggl.
     \Pi_{Q^\ast Q^\ast}^{(a)[\sigma Q^\ast]}
     - \Pi_{PP}^{(d)[\sigma Q]}
   \biggr\vert_{\rm div}
  = \frac{2k^2}{F_\sigma^2}
    \left[
      - \frac{\Lambda^3}{24\sqrt{2}\pi^2}
      + \frac{M_\rho^2}{4\pi^2}
        \left( \frac{\Lambda}{\sqrt{2}} - M_G \ln \Lambda \right)
    \right]
\ ,
\nonumber\\
&& \biggl.
     \Pi_{Q^\ast Q^\ast}^{(d)[\sigma P^\ast]}
     - \Pi_{PP}^{(a)[\sigma P]}
   \biggr\vert_{\rm div}
  = \frac{2k^2}{F_\sigma^2}
    \left[
      \frac{\Lambda^3}{24\sqrt{2}\pi^2}
      - \frac{M_\rho^2}{4\pi^2}
        \left( \frac{\Lambda}{\sqrt{2}} - M_H \ln \Lambda \right)
    \right]
\ .
\end{eqnarray}
These lead to
\begin{equation}
\biggl. \Pi_{Q^\ast Q^\ast}^{(a+d)}
  - \Pi_{PP}^{(a+d)} \biggr\vert_{\rm div}
  = - g^2 \frac{k^2}{2\pi^2} ( M_G - M_H ) \ln \Lambda
\ ,
\label{QQPP ad}
\end{equation}
where we used $M_\rho^2 = g^2 F_\sigma^2$. The remaining
contributions sum to
\begin{equation}
\biggl. \Pi_{Q^\ast Q^\ast}^{(c+f)}
  - \Pi_{PP}^{(c+f)} \biggr\vert_{\rm div}
  = - g^2 \frac{1 - 2k - 2k^2 }{2\pi^2} ( M_G - M_H ) \ln \Lambda
\ .
\label{QQPP cf}
\end{equation}

By summing up the contributions in Eqs.~(\ref{QQPP be}),
(\ref{QQPP ad}) and (\ref{QQPP cf}),  we obtain the divergent part
of the correction to the mass difference:
\begin{equation}
\biggl. \Pi_{Q^\ast Q^\ast} - \Pi_{PP} \biggr\vert_{\rm div}
 = - {\mathcal C}_2(N_f)\,
   \frac{g^2}{2\pi^2}\Bigl( 1 - 2k - k^2 \Bigr)
   (M_G - M_H)\ln\Lambda
\ ,
\label{quantum correction app}
\end{equation}
where we reinstated the group factor ${\mathcal C}_2(N_f)$. The
logarithmic divergence in the above expression is renormalized by
the bare contribution given by
\begin{equation}
\biggl. \Pi_{Q^\ast Q^\ast} - \Pi_{PP} \biggr\vert_{\rm bare}
  = \Delta M_{\rm bare} \ .
\end{equation}
Thus the RGE for the mass difference
$\Delta M = M_G - M_H$ has the following form:
\begin{equation}
 \mu\frac{d\,\Delta M}{d\mu}
 =  {\mathcal C}_2(N_f)\,
   \frac{g^2}{2\pi^2}\Bigl( 1 - 2k - k^2 \Bigr)\Delta M.
\label{rge}
\end{equation}
We should stress that this RGE is exactly the same as the one
in Eq.~(\ref{rge a=1}) obtained by setting $a=1$, i.e., $F_\sigma = F_\pi$.

\section{Need for a precise value of the quark condensate}
\label{quark condensate}

We emphasized in the main text that there
is a great deal of uncertainty on the value of the quark
condensate at the relevant matching scale $\Lambda_M$. In this
Appendix, we list a few examples to show what sort of uncertainty
we are faced with.

We took $\left\langle \bar{q} q \right\rangle_{\rm 1\,GeV}
=-(225\pm25\,\mbox{MeV})^3$ in our analysis as a ``standard
value." For comparison, we shall take two other values quoted in
Ref.~\cite{BL} (without making any judgments on their validity).
Consider therefore
\begin{eqnarray}
\left\langle \bar{q} q \right\rangle_{\rm 1\,GeV}
&=& -(225\pm25\,\mbox{MeV})^3 \ ,
\nonumber\\
\left\langle \bar{q} q \right\rangle_{\rm 2\,GeV}
&=&
\left\{\begin{array}{l}
 -(273\pm19\,\mbox{MeV})^3 \ , \\
 -(316\pm24\,\mbox{MeV})^3 \ .
\end{array}\right.
\end{eqnarray}
Brought by RGE to the scale we are working at, $\Lambda_M=1.1$
GeV, and substituted into our formula for $\Delta M$, we get the
corresponding quantum corrected splitting
\begin{eqnarray}
\Delta M = \left\{\begin{array}{l}
 0.31\pm 0.12\,\mbox{GeV} \ , \\
 0.43\pm 0.12\,\mbox{GeV} \ , \\
 0.67\pm0.20\,\mbox{GeV} \ .
\end{array}\right.
\end{eqnarray}
This result clearly shows that the splitting cannot be pinned down
unless one has a confirmed quark condensate.

%%%%%%%%%%%%%%%%%%%%%%%%%%%%%%%%%%%%%%%%%%%%%%%%%%%%%%%%%%%%%%%%%%%%
%%%%%%%%%%%%%%%%%%%%%%%%%%%%%%%%%%%%%%%%%%%%%%%%%%%%%%%%%%%%%%%%%%%%
\pagebreak

\end{document}